\documentclass[aip,apl,amsmath,amssymb,reprint]{revtex4-1}
\usepackage{graphicx}
\usepackage{times}

\usepackage{color}

\usepackage{multirow}
\usepackage{makecell}

\begin{document}

\title{Ultrafast sensing of photoconductivity decay using microwave resonators}

\author{B. Gy\"{u}re-Garami}
\affiliation{Department of Physics, Budapest University of Technology and Economics and MTA-BME Lend\"{u}let Spintronics Research Group (PROSPIN), Po. Box 91, H-1521 Budapest, Hungary}

\author{B. Blum}
\affiliation{Department of Physics, Budapest University of Technology and Economics and MTA-BME Lend\"{u}let Spintronics Research Group (PROSPIN), Po. Box 91, H-1521 Budapest, Hungary}

\author{O. S\'agi}
\affiliation{Department of Physics, Budapest University of Technology and Economics and MTA-BME Lend\"{u}let Spintronics Research Group (PROSPIN), Po. Box 91, H-1521 Budapest, Hungary}

\author{A. Bojtor}
\affiliation{Department of Physics, Budapest University of Technology and Economics and MTA-BME Lend\"{u}let Spintronics Research Group (PROSPIN), Po. Box 91, H-1521 Budapest, Hungary}

\author{S. Kollarics}
\affiliation{Department of Physics, Budapest University of Technology and Economics and MTA-BME Lend\"{u}let Spintronics Research Group (PROSPIN), Po. Box 91, H-1521 Budapest, Hungary}

\author{G. Cs\H{o}sz}
\affiliation{Department of Physics, Budapest University of Technology and Economics and MTA-BME Lend\"{u}let Spintronics Research Group (PROSPIN), Po. Box 91, H-1521 Budapest, Hungary}

\author{B. G. M\'{a}rkus}
\affiliation{Department of Physics, Budapest University of Technology and Economics and MTA-BME Lend\"{u}let Spintronics Research Group (PROSPIN), Po. Box 91, H-1521 Budapest, Hungary}

\author{J. Volk}
\affiliation{Institute of Technical Physics and Materials Science, Centre for Energy Research, Konkoly-Thege M. \'{u}t 29-33, H-1121 Budapest, Hungary}

\author{F. Simon}
\email[Corresponding author: ]{f.simon@eik.bme.hu}
\affiliation{Department of Physics, Budapest University of Technology and Economics and MTA-BME Lend\"{u}let Spintronics Research Group (PROSPIN), Po. Box 91, H-1521 Budapest, Hungary}

\date{\today}

\begin{abstract}
Microwave reflectance probed photoconductivity (or $\mu$-PCD) measurement represents a contactless and non-invasive method to characterize impurity content in semiconductors. Major drawbacks of the method include a difficult separation of reflectance due to dielectric and conduction effects and that the $\mu$-PCD signal is prohibitively weak for highly conducting samples. Both of these limitations could be tackled with the use of microwave resonators due to the well-known sensitivity of resonator parameters to minute changes in the material properties combined with a null measurement. A general misconception is that time resolution of resonator measurements is limited beyond their bandwidth by the readout electronics response time. While it is true for conventional resonator measurements, such as those employing a frequency sweep, we present a time-resolved resonator parameter readout method which overcomes these limitations and allows measurement of complex material parameters and to enhance $\mu$-PCD signals with the ultimate time resolution limit being the resonator time constant. This is achieved by detecting the transient response of microwave resonators on the timescale of a few 100 ns \emph{during} the $\mu$-PCD decay signal. The method employs a high-stability oscillator working with a fixed frequency which results in a stable and highly accurate measurement. 
\end{abstract}
\maketitle

\section{Introduction}

Microwave detected photoconductivity measurement, $\mu$-PCD \cite{Ohsawa1983,Kunst1986,muPCD_summary1}, is a standard laboratory and industrial tool to characterize the amount of light-excited charge carriers and their lifetime. Knowledge of these parameters is crucial for semiconductor applications in light harvesting and also for the characterization of impurity concentrations. Examples include the characterization of Co and Fe impurities in silicon wafers \cite{muPCD_summary1,muPCD_Co_characterization}, the light-induced carrier recombination rate measurements in novel perovskite based photovoltaic materials \cite{Lami_muPCD_Chouhan,Lami_muPCD_Labram,Lami_muPCD_Guse,Lami_muPCD_Bi}, non-silicon semiconductors, e.g. CdSe and CdTe (Ref.~\onlinecite{Novikov_CdTe}) and novel low-dimensional materials including carbon nanotubes \cite{FreitagNL2003,LuNT2006}, graphene \cite{VaskoPRB2008,DochertyNatCom2012}, transition metal dichalcogenides \cite{XiaNatPhot2014}, and black phosphorus \cite{LiuAdvMat2016,MiaoACSNano2017}.

The most conventional $\mu$-PCD implementations rely on detecting reflection of microwaves from a sample which is irradiated using an antenna or an open waveguide, i.e. by non-resonant means. Besides its simplicity, this approach usually suffers from a large, non light-induced reflected background signal that can either saturate the receiver electronics (thus preventing measurement of small light-induced reflections) or perplex the phase sensitive analysis of the back-reflected microwave signal. Measurement of the phase is required in order to separate the reflection due to dielectric and conduction effects. This hindrance is especially pronounced for samples with a high conductivity, i.e. for a doped semiconductor.

It was recognized back in 1977 (Refs.~\onlinecite{Warman_RadPhysChem_1977,fessenden1982_AFC_muPCD_meas}) and reiterated recently (Ref.~\onlinecite{Reid_2017}) that the use of microwave resonators could eliminate the above mentioned problems since resonators allow for a \emph{null measurement} and also to improve the sensitivity of the method due to the well-known amplifying effect of resonators. The latter can be best demonstrated for considering a single resistor with resistance $R$ whose value changes by $\Delta R$. In case the resistor is part of a high quality factor ($Q$) RLC circuit, whose impedance is matched to a waveguide with wave-impedance of $Z_0$ ($Z_0 \gg R$), the change of the return impedance near the resonance frequency of the resonator is $\Delta Z=\Delta R\cdot Z_0/R$ (Ref.~\onlinecite{PozarBook}), i.e. the sensitivity to a small resistance change is significantly enhanced (we present a lumped element circuit model calculation in the Supplementary Materials). Equivalently, it is often expressed as $\Delta R/R=\Delta Q/Q$ (Ref.~\onlinecite{cetinoneri2010_time_resolved_muPCD}).

The known approach to combine the $\mu$-PCD measurement with microwave resonators involves the detection of power reflected from a resonator during and after a light pulse \cite{SubramanianJAP1998,Novikov_CdTe}. However, this method ignores the information contained in the phase (due to the power detection). In addition, both frequency and $Q$-factor changes lead to a modified reflection, thus modeling to obtain material dependent parameters is rather involved \cite{Amato,fessenden1982_AFC_muPCD_meas,Eckstrom}. Nevertheless, the approach yields effective $\mu$-PCD lifetime values and it was successful in constructing sensitive electromagnetic radiation detectors \cite{Tepper_detector_2000,Tepper_detector_2001,cetinoneri2010_time_resolved_muPCD,Braggio_detector_2014}.

Clearly, a modeling-independent, time-resolved measurement of resonator $Q$ and eigenfrequency, $f_0$, is highly desired.
It is well known that the finite bandwidth of resonators inherently limits time-resolved measurements to the resonator time constant of
$\tau=2Q/\omega_0$, with typical values of $\tau=3\dots 300\,\text{ns}$ for $Q=100\dots 10,000$ and $f_0=10\,\text{GHz}$. However, an additional misconception is that the practical measurement of the $Q$ factor and $f_0$ is limited further to a few ms or even seconds, depending on the analyzing electronics in a frequency swept experiment using e.g. a network analyzer \cite{PetersanAnlage,LuitenReview,KajfezReview}. This is prohibitively slow for a meaningful time-resolved resonator measurement. 

We recently reported a novel method \cite{GyureRSI,GyureGaramiRSI} which allows measurement of the resonator parameters, $f_0$ and $Q$, with a rapid time domain measurement. The technique shows great resemblance to pulsed nuclear magnetic resonance methods \cite{Ernst} and to the optical cavity ring-down spectroscopy \cite{optics1,optics2,optics3,optics4}. The pulsed resonator readout method employs a short RF pulse with carrier frequency not necessarily matching $f_0$ ($f_{\text{LO}}\neq f_0$) which induces both a switch-on and switch-off transients. Both transients represent a decaying oscillation on the eigenfrequency of the resonator, $f_0$ (Ref.~\onlinecite{SchmittZimmer}) and with its time constant, $\tau$ (Refs.~\onlinecite{Gallagher,KomachiTanaka,Amato,EatonTransient}). The transient signals can be conveniently downconverted with the same $f_{\text{LO}}$ frequency as the exciting signal. Such measurements in the time domain have two important advantages: enhanced accuracy (also known as the Connes advantage \cite{Connes}) since the measurement is traced back to a stable frequency, and a simultaneous measurement of the whole resonator response (also known as Fellgett or multiplex advantage \cite{Fellgett}). 

This time-resolved, pulsed resonator readout method has been successfully employed to evidence a heating related microwave absorption anomaly in carbon nanotubes \cite{MarkusGyurePSSB} and to improve the measurement accuracy of power absorbed from an RF field in magnetic ferrite nanoparticles during hyperthermia \cite{GresitsJPD}. These results motivate the present study to explore the possibility to use this method for the detection of time-resolved $\mu$-PCD studies in silicon. Herein, we report time-resolved $\mu$-PCD measurements for a silicon single crystal samples which are placed inside a microwave cavity resonator with a time constant of about 100 ns. A Q-switch laser induces extra electron-hole pairs in the sample and the laser repetition is synchronized with a train of short microwave pulses. The resulting switch-off transient is detected after each microwave pulse, which yields information about the sample photoconductivity. We present time-resolved resonator $f_0$ and $Q$ data on a silicon wafer sample with $\mu$-PCD lifetime around 100 $\mu$s. The measurement clearly demonstrates the utility of the pulsed resonator readout method for the detection of time-resolved microwave detected photoconductivity measurements.


\section{Principle of the $\mu$-PCD measurements}

\subsection{The conventional $\mu$-PCD method}

In order to illustrate the advances of the present method, we recapture the principle of the conventional $\mu$-PCD studies. The conventional $\mu$-PCD method is based on detecting the reflection of microwaves from a semiconductor wafer. Reflection of electromagnetic waves from a material can be most conveniently described with the introduction of the wave impedance of the material\cite{PozarBook,chen2004microwave}: 
\begin{gather}
Z=\sqrt{\frac{i\omega \mu}{\sigma+i \omega \epsilon}}
\label{wave_impedance}
\end{gather}
where $\mu=\mu_0\mu_{\text{r}}$ is the permeability, $\epsilon=\epsilon_0\epsilon_{\text{r}}$ is the permittivity, $\omega$ is the angular frequency of the wave, and $\sigma$ is the conductivity of the sample. In general, $\mu$, $\epsilon$, and $\sigma$ are complex quantities. Note that Eq.~\eqref{wave_impedance} returns the well-known wave impedance of vacuum $Z_0=\sqrt{\mu_0/\epsilon_0}\approx 377\,\Omega$ for $\mu_{\text{r}}=\epsilon_{\text{r}}=1$ and $\sigma=0$. For conductors, $Z$ is denoted by the surface impedance, $Z_{\text{s}}$, which highlights that electromagnetic waves penetrate only into the surface of metals. Then ($\sigma$ finite, $\mu_{\text{r}}=\epsilon_{\text{r}}=1$) and within the quasi-stationary approximation ($\sigma\gg\omega\epsilon$), the above formula for $Z$ returns the well-known expression for the surface impedance  of $Z_{\text{s}}=\sqrt{\frac{i \omega \mu_0}{\sigma}}=\frac{1+i}{2}\mu_0\omega\delta$, where the penetration depth reads: $\delta=\sqrt{\frac{2}{\mu_0\omega \sigma}}$. It also implicitly contains the often used complex dielectric constant for a lossy dielectric (i.e. a dielectric with a finite $\sigma$): $\widetilde{\epsilon_{\text{r}}}=\epsilon_{\text{r}}-\frac{i \sigma}{\epsilon_0 \omega}$ ($\epsilon_{\text{r}}$ is the real dielectric constant of the material).

We then consider that the material occupies the half space and it has a surface impedance of $Z_{\text{s}}$. An electromagnetic wave interacts with it, which propagates in free space or in a generic waveguide (it could be a coaxial cable or a TE10 microwave waveguide) thus this medium is modeled with a wave impedance of $Z_{\text{wg}}$. The radiation is reflected from the material and the reflection coefficient (i.e. ratio of the transmitted to reflected field voltage or amplitude), $\Gamma$, is given as \cite{PozarBook}:

\begin{gather}
\Gamma=\frac{Z_{\text{s}}-Z_{\text{wg}}}{Z_{\text{s}}+Z_{\text{wg}}}.
\label{Fresnel_formula}
\end{gather}
In the so-called $S$ parameter representation, $S_{11}\approx\Gamma$ holds (the equality is strictly valid for zero transmission, see Ref.~\onlinecite{chen2004microwave}, p. 196). For a $\mu$-PCD experiment in an industrial environment, the reflection is often detected by an antenna, which induces additional geometric factors but it does not affect the generic physical description in Eq.~\eqref{Fresnel_formula}.

Eq.~\eqref{Fresnel_formula} is related to the usual Fresnel reflection formula of $r=\frac{1-n}{1+n}$ (for normal incidence of electromagnetic waves from vacuum, $n=1$) by recognizing the relation between the wave impedance and the index of refraction as: $Z=Z_0/\widetilde{n}$. The complex index of refraction for a non-magnetic material reads $\widetilde{n}=\sqrt{\widetilde{\epsilon_{\text{r}}}}=\sqrt{\epsilon_{\text{r}}-\frac{i \sigma}{\epsilon_0 \omega}}$.

Alternatively for good conductors, the reflection at low frequencies can be approximated with the Hagen-Rubens relation: 
\begin{gather}
R=\left|r\right|^2=1-2\sqrt{\frac{2\epsilon_0\omega}{\sigma}}.
\end{gather}

\begin{figure}[htp]
\begin{center}
\includegraphics[width=0.45\textwidth]{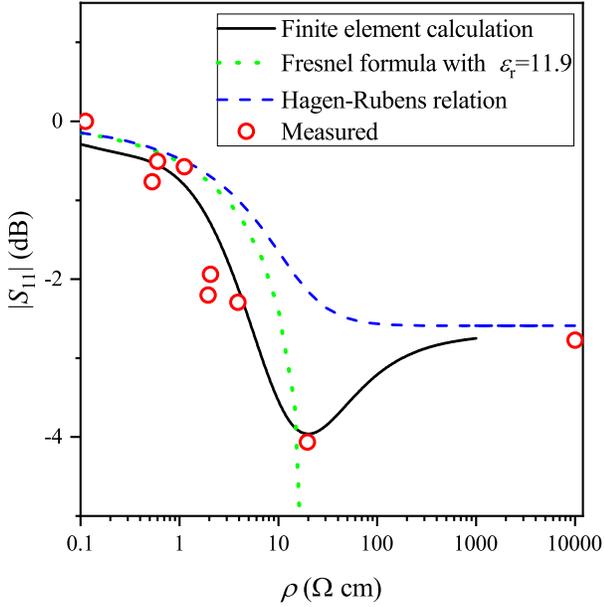}
\caption{Reflected microwave amplitude $|S_{11}|$ as a function of the silicon wafer resistivity. Solid curve is a finite element calculation for a WR90 X-band waveguide, covered with the wafers, dashed curve is calculated using the Fresnel formula (as explained in the text), dotted curve is calculation with the Hagen-Rubens relation, and symbols show the experimental data for a set of silicon single crystal wafers.}
\label{Fig:WG_Reflection}
\end{center}
\end{figure}

The reflection amplitude, $S_{11}$, from a silicon single crystal wafer with varying resistivity is shown in Fig.~\ref{Fig:WG_Reflection} for various approximations: a finite element electromagnetic modeling for the wafer covering a WR90 X-band waveguide (TE10 mode, 8-12.4 GHz), the Fresnel formula, the Hagen-Rubens relation. We found that the calculated reflections are little affected when we considered the small variation of $\epsilon_{\text{r}}$ as a function of doping, according to Ref.~\onlinecite{EpsilonVariation}. The figure demonstrates that the Hagen-Rubens relation falls well onto the other two calculations in its domain of validity. 

Fig.~\ref{Fig:WG_Reflection}. also contains the experimental results. To obtain these, we covered the WR90 X-band waveguide with a series of silicon single crystal wafers with varying resistivity from $\varrho=0.528\,\Omega\cdot\text{cm}$ up to $\varrho=10\,\text{k}\Omega\cdot\text{cm}$, i.e. through 4 orders of magnitude in $\varrho$. We used a copper plate covering the waveguide as $|S_{11}|=0\,\text{dB}$ reference. 
As expected, the experimental data lies close to the result of the finite element electromagnetic modeling. The Fresnel formula also demonstrates well the general trend in the reflected amplitude, even if it deviates from the electromagnetic modelling.

In principle, the reflection approach allows to determine the real and imaginary parts of the material parameters from the phase sensitive detection of the reflected microwaves. However, this measurement requires an accurate calibration of the reflected microwave phase. In addition, most $\mu$-PCD measurements, which are implemented in an industrial environment, measure the reflected microwave power only. In contrast, as we shall show below, a measurement of the material parameters in a microwave resonator allows for the automatic disentanglement of the real and imaginary parts of the material parameters.

Nevertheless, the major hindrance of the conventional $\mu$-PCD method is that a substantial reflection is present already in dark conditions: as Fig.~\ref{Fig:WG_Reflection}. shows, for most cases the reflection is around 3 dB, i.e. half of the microwave power is reflected even without illumination. It clearly hinders the detection of the extra, light-induced reflection by the saturation of the detecting electronics and the always present dark background gives rise to additional shot noise.  

\subsection{The resonator based $\mu$-PCD method}

The so-called cavity perturbation method \cite{buravov71,Gruner1} is applicable for a sample which is placed inside a microwave cavity resonator. The presence of the sample affects both the resonance frequency, $f_0$, and quality factor, $Q_0$, of the unloaded resonator. It was derived in Ref.~\onlinecite{LandauBook} that the resonator perturbation for a cylinder with diameter $a$ reads:

\begin{gather}
\frac{\Delta f}{f_0}-i\Delta\left(\frac{1}{2Q}\right)=-\gamma \alpha
\label{Landau_cavity_perturbation}
\end{gather}
where $\gamma$ is a sample size dependent constant (also depends on the cavity mode and electromagnetic field distribution). $\Delta f$ is the shift in the resonant frequency and $\Delta\left(\frac{1}{2Q}\right)$ is the additional, sample related loss in the cavity. The authors of Ref.~\onlinecite{LandauBook} introduced the $\alpha$ polarizability:

\begin{gather}
\alpha=-2\left(1-\frac{2}{a\widetilde{k}}\frac{J_1\left(a\widetilde{k} \right)}{J_0\left(a\widetilde{k} \right)} \right)
\label{Landau_polarizability}
\end{gather}
with $\widetilde{k}=i \omega \sqrt{\mu \epsilon}\sqrt{1-\frac{i}{\omega\epsilon\varrho}}$ being the complex wavenumber of the microwaves inside the material. $J_0$ and $J_1$ are Bessel functions of the first kind.

In the limit of finite electromagnetic wave penetration into the sample, Eq.~\eqref{Landau_cavity_perturbation} reduces to the better known expression which relates the resonator parameters directly to the surface impedance according to Eq.  \eqref{wave_impedance}, as follows \cite{PozarBook,chen2004microwave,Gruner1,Gruner2,Gruner3}:

\begin{gather}
\frac{\Delta f}{f_0}-i\Delta\left(\frac{1}{2Q}\right)=-i \nu Z_{\text{s}}
\label{surface_imp_cavity_perturbation}
\end{gather}
where $\nu$ is a geometry factor (not dimensionless) that is proportional to the ratio of the sample surface to the cavity surface but it also depends on the resonator mode. We discuss an additional sample geometry and explicitly derive the relation between Eqs.~\eqref{Landau_cavity_perturbation} and \eqref{surface_imp_cavity_perturbation} in the Supplementary Materials. 

Eq.~\eqref{Landau_cavity_perturbation} shows that measurement of the cavity frequency shift and loss allows to disentangle the real and imaginary parts of the material wave impedance. A limitation of the method is that the geometry factor is generally unknown therefore a calibrating measurement is required to obtain absolute material parameter values. The resonator based method prevails when the relative changes in the material parameter is required as a function of time.

\begin{figure}[htp]
\begin{center}
\includegraphics[width=0.45\textwidth]{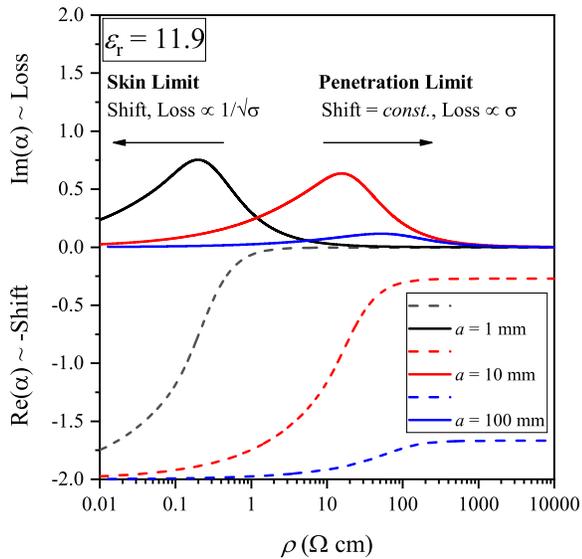}
\caption{Variation of the resonator parameters according to Eq.~\eqref{Landau_cavity_perturbation} with varying silicon resistivity. The two limiting cases are indicated, when microwaves are limited to the skin depth only (skin limit) and when they can penetrate into the sample (penetration limit). Note the characteristically different behavior of the sample parameters in the two regimes versus the sample resistivity.}
\label{Fig:Resonator_reflection}
\end{center}
\end{figure}

Fig.~\ref{Fig:Resonator_reflection}. summarizes the change of a microwave resonator parameters for a sample with varying resistivity according to Eq.~\eqref{Landau_cavity_perturbation} with the $\epsilon_{\text{r}}=11.9$ for silicon. The behavior can be split to two regimes depending on whether the microwaves penetrate into the sample (penetration limit) or whether it is limited by the skin-effect. For the earlier, the shift is constant and the loss, $\Delta(1/2Q)$, is linear to $\sigma$. In the latter, the skin limit, the real and imaginary parts of $Z_{\text{s}}$ are equal and are both proportional to $1/\sqrt{\sigma}$. This correspondence allows to obtain the material parameters from the measurement of the cavity, besides the $\nu$ geometry factor. However, the major advantage of using the microwave resonators is the essentially null measurement it provides.

We emphasize that Eq.~\eqref{Landau_cavity_perturbation} gives the cavity perturbation formula for an arbitrary value of $\sigma$ and $\epsilon_{\text{r}}$. Often one discusses the two extremal cases for the cavity perturbation only: e.g. for $\mu$-PCD studies on gas or liquid plasmas \cite{Warman_RadPhysChem_1977} or on materials with a low conductivity \cite{SubramanianJAP1998} the penetration limit is discussed only, whereas the skin-limit with the surface impedance description is used for good conductors \cite{HolczerPRB}.
While the full analysis of $\sigma$ and $\epsilon_{\text{r}}$ can be performed for the case of cavity perturbation, this is beyond the scope of the present contribution and we focus on the technical development, i.e. on the time-resolved measurement of the resonator shift and loss.

\section{The resonator based photoconductivity measurement}

\subsection{The measurement setup}

\begin{figure}[htp]
\begin{center}
\includegraphics[width=0.45\textwidth]{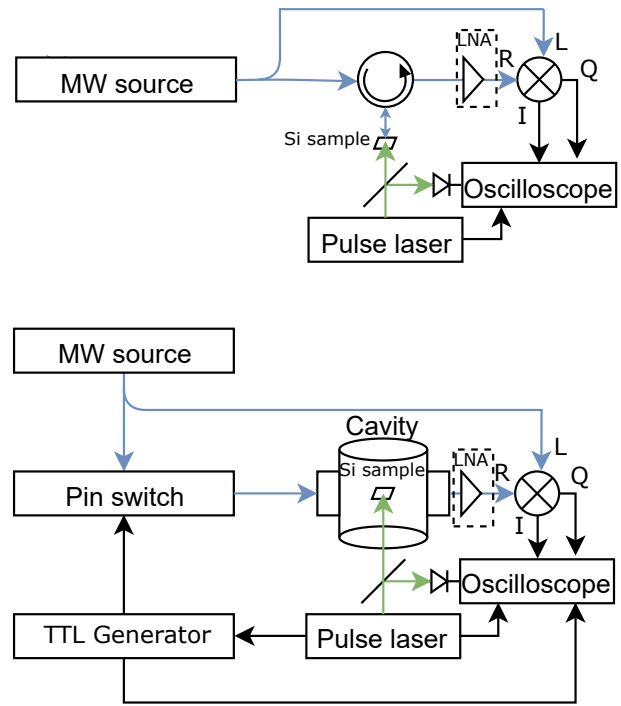}
\caption{Schematics of the conventional (upper panel) and resonator based $\mu$-PCD decay experiments. A Q-switch laser provides light excitation in both cases. The microwaves are detected with an IQ mixer. In the conventional setup, the signal is measured using reflectometry, whereas in the novel approach, it is measured through a microwave cavity. A microwave IQ mixer detects the signal in both cases and an optional LNA is indicated with a dashed box. A number of microwave isolators are not shown in the figure.}
\label{Fig:ExpSetup}
\end{center}
\end{figure}

Our setup for the time-resolved $\mu$-PCD measurement is shown in Fig.~\ref{Fig:ExpSetup} including both the conventional (upper panel) and the novel, resonator based approach (lower panel). A Q-switch pulse laser (527 nm Coherent Evolution-15, Nd:YLF) with 1 kHz
repetition frequency and $\sim 300\,\text{ns}$ pulse duration is used for the excitation of charge carriers in the
semiconductor samples.  We note that the 527 nm excitation is capable of photoexciting charge carriers in silicon, even though its band edge is around 1100 nm, which would be a more efficient wavelength for such purposes.

The microwave source is a PLL locked synthesizer (HP-Agilent 83751B or a K\"uhne Electronic GmbH model MKU LO 8-13 PLL) which drives the LO of an IQ mixer (Marki Microwave IQ0618LXP double-balanced mixer, LO/RF: 6-18 GHz, IF: DC-500 MHz, 7.5 dB conversion loss). The mixer downconverts the incoming RF signal and the I and Q signals are digitized with an oscilloscope (Tektronix MDO-3024, BW=200 MHz). 

Optionally, the RF signal can be amplified by a low noise amplifier (LNA, JaniLab Inc., NF=1.4 dB, Gain=15 dB, 1 dB compression point, P1dB, 10 dBm), which is indicated by a dashed box in the figure. Both the LO and RF inputs of the mixer are isolated galvanically from the rest of the circuit with band-pass (8-12 GHz) DC-blocks. The rising edge of the laser pulses are detected with a fast photodiode (Thorlabs DET36A/M) which provides a jitter-free trigger signal.

This signal triggers the oscilloscope in the conventional setup: therein a standard X-band (8-12.4 GHz) WR90 waveguide is used to irradiate samples. The silicon wafers fully cover the waveguide and are illuminated by the light, whose beam aperture is such that it roughly covers the entire waveguide opening. We checked that the laser illumination from the front (i.e. opposite to the microwave irradiation direction) gives qualitatively identical results to those when the sample is irradiated from the back (i.e. parallel to the microwave irradiation direction). The only difference is that irradiation from the back results in smaller signals as the microwave waveguide limits the insertion of light. A standard X-band circulator (Ditom Microwave Inc.) acts as duplexing unit between the exciting and reflected microwaves.

In the novel setup, the sample is inside a microwave cavity resonator operating in the TE011 mode (with an unloaded quality factor $Q_0\approx 5000$) and we use it in  transmission. The cavity is undercoupled for both the input and output ($\beta_{\text{input}} \approx \beta_{\text{output}} \approx 1/3$) which represents a compromise between the resonator bandwidth and transmitted signal \cite{PozarBook}. The parameters of the resonator are measured with the transient method \cite{GyureRSI,GyureGaramiRSI}: the exciting microwaves are pulsed, which forces the cavity to transmit microwaves in a transient state. Although the exciting carrier frequency, $f_{\text{LO}}$, does not necessarily match the resonator eigenfrequency, still the transient signal oscillates on the resonant frequency of the cavity, $f_0$. The carrier of the excitation frequency, $f_{\text{LO}}$, is intentionally detuned from $f_0$ in order to detect the transient with an intermediate frequency around $5\dots 10$ MHz, which removes the 1/$f$ noise of the mixer.

The microwave pulses are formed with a fast PIN diode switch (Advanced Technical Materials, S1517D, 5 ns 10-90\% rise-fall transient) which is driven by a TTL signal. This signal contains a switch-on of 0.5 $\mu$s and is repeated every 2 $\mu$s. This duration and repetition are well suited for our cavity with $\tau \approx 100\,\text{ns}$ but these could be further reduced for a cavity with a lower $Q$, which would allow for the detection of even faster transients. The optical trigger provides the synchronizing signal for an arbitrary waveform generator (Siglent SDG1025) which generates a train of TTL pulses.

\subsection{Resonator transient measurements}

\begin{figure}[htp]
\begin{center}
\includegraphics[width=0.45\textwidth]{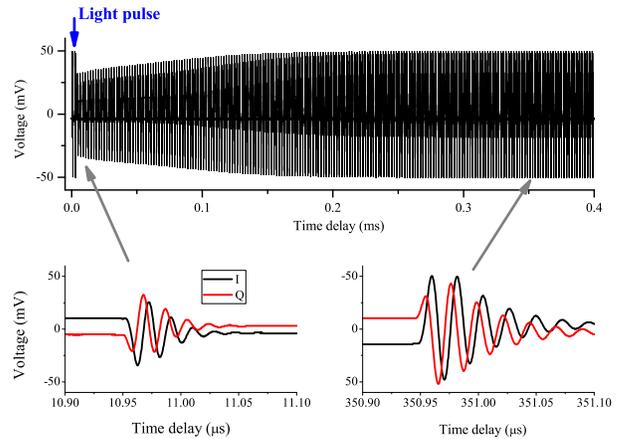}
\caption{Scheme of the cavity transient detected $\mu$-PCD. The resonator transients appear as a train of signals, which contain transients with different frequency and linewidth depending on the state of the sample (\textit{Upper panel.}). Two examples for such quadrature detected traces (I and Q signals) are shown for different delay times after the light pulse (\textit{Lower panel.}). These traces are Fourier transformed to yield the microwave cavity resonance curves.}
\label{Fig:measurement_scheme}
\end{center}
\end{figure}

An example for the time-resolved microwave cavity transient method is depicted in Fig.~\ref{Fig:measurement_scheme}. The Q-switch laser pulse (1 ms repetition rate) triggers a train of pulses (each with a duration of 0.5 $\mu$s followed by another 1.5 $\mu$s waiting time) which drives the microwave PIN diode. The microwave cavity responds with switch-on and off transients. We measure the microwave transients immediately after switching off the microwave excitation as therein the exciting microwave signal is absent. Thus the transient contains information about the resonator only, free from any further signals and can thus be considered as a \emph{null measurement} of the relevant information. 

Two examples for such IQ traces are shown in Fig.~\ref{Fig:measurement_scheme} for different time delays after the light pulse. These signals are then Fourier transformed to which Lorentzian curves can be fitted. The fitting yields the eigenfrequency and bandwidth of the cavity as a function of the time delay. These directly give the microwave resonator shift and loss, which allows determination of the material parameters according to Eq. \onlinecite{LandauBook}.

\begin{figure}[htp]
\begin{center}
\includegraphics[width=0.5\textwidth]{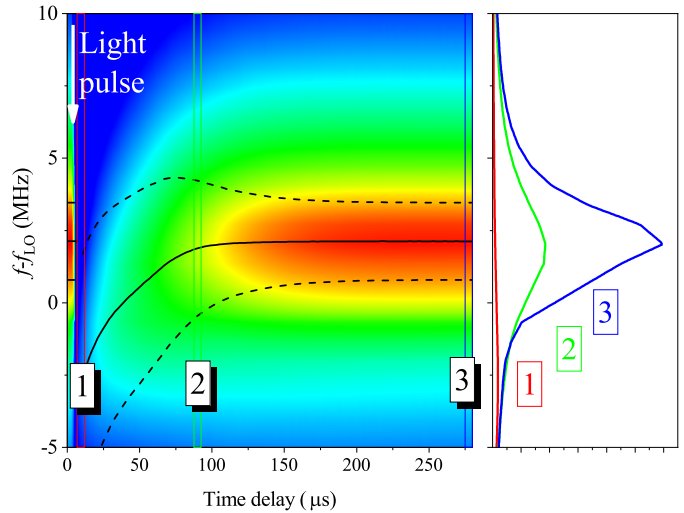}
\caption{Time-resolved microwave cavity detected $\mu$-PCD traces for a silicon sample ($\varrho=19.7\,\Omega\,\text{cm}$). The contour plot was obtained by recording consecutive cavity transients after a switch on duration of 0.5 $\mu$s and a repetition time of 2 $\mu$s. The contour plot has a logarithmic scale to better show the smaller trace values. Solid curve is the shifting of the resonator $f_0$ with respect to the LO frequency and dashed curves indicate the value of the half width of the Lorentzian. The vertical separation between the two dashed curves is the resonator bandwidth. The profiles on the right hand side are from the indicated time positions.}
\label{Fig:muPCD_Demonstration}
\end{center}
\end{figure}

This type of measurement can be also conveniently shown in a three-dimensional contour plot. In Fig.~\ref{Fig:muPCD_Demonstration}., we show the result of the time-resolved resonator readout method for a single crystal silicon wafer sample ($\varrho=19.7\,\Omega\,\text{cm}$) with a relatively long (about 100 $\mu$s) charge carrier recombination time.  The contour plot also shows the time-dependent $f_0-f_{\text{LO}}$ (solid curve) and the half maximum value points of the Lorentzian (dashed curves). The vertical separation between the latter two curves is the resonator bandwidth, BW, which gives $Q=f_0/\text{BW}$. A clear time dependence of both $f_0$ and $Q$ is observable from the data. The right hand side of Fig.~\ref{Fig:muPCD_Demonstration}. shows individual Lorentzian resonance profiles which are shown for three different time delays. 

\begin{figure}[htp]
\begin{center}
\includegraphics[width=0.45\textwidth]{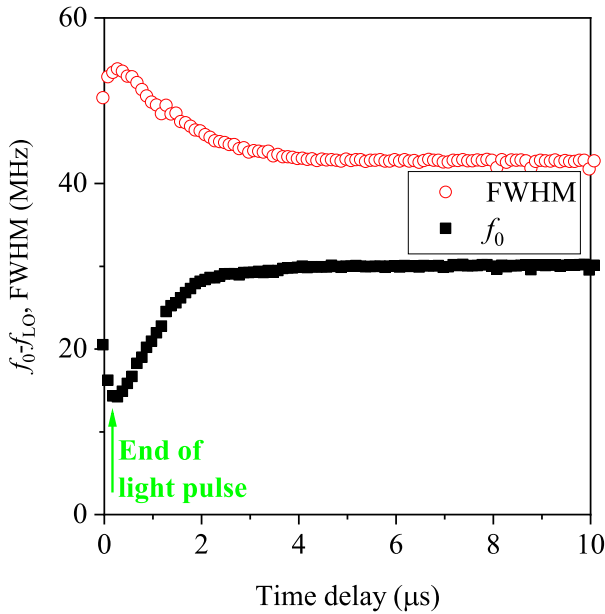}
\caption{Demonstration of the time-resolved microwave cavity detected $\mu$-PCD traces for an ultrafast case. The sample is a silicon wafer ($\varrho=0.5\,\Omega\,\text{cm}$). Each individual $f_0-f_{\text{LO}}$ and FWHM data points were obtained from consecutive cavity transients containing a switch on duration of 50 ns and a repetition time of 200 ns. The latter time resolution, $\Delta t$ is indicated by an arrow (not to scale). Note that the cavity is strongly loaded with this sample, thus $Q\approx 250$.}
\label{Fig:muPCD_Demonstration_fast}
\end{center}
\end{figure}

Fig.~\ref{Fig:muPCD_Demonstration_fast} shows a time-resolved resonator detected $\mu$-PCD traces for a Si wafer sample which showed an ultrafast charge carrier dynamics less than 2 $\mu$s. This was performed on a sample with an already low resistivity, $\varrho=0.5\,\Omega\,\text{cm}$, which reduced the cavity quality factor to about $Q\approx 250$. This results in a short cavity transient time of about $\tau=8\,\text{ns}$. This allowed to perform the cavity transient experiment with a repetition time of 200 ns (time resolution is about the symbol size in the figure), which contained a switch on duration of 50 ns. Clearly, the time-dependent variation of both $f_0$ and the BW ($Q$) can be observed from the data. This shows that our method works well for charge carrier life-times down to the microsecond range. 

Finally, we highlight several key points of the present development: our approach does not require frequency stabilization, or AFC, which was required in alternative studies \cite{SubramanianJAP1998,fessenden1982_AFC_muPCD_meas}, except that the irradiating microwave pulse should be within about 10-100 times the resonator BW with respect to the resonance frequency. Another important aspect is that we obtain the resonator parameters, $f_0$ and $Q$, \emph{directly} from the data, without the need for an involved modeling of the microwave cavity transmission or reflection. Nevertheless, obtaining the time-dependent material parameters ($\sigma$ and $\epsilon_{\text{r}}$) also requires a calculation according to Eq.~\eqref{Landau_cavity_perturbation}.

The utility of the present method in an industrial environment remains to be addressed. We believe that it may find better applications in the research of novel semiconductors such as e.g. novel photovoltaic perovskites \cite{Lami_muPCD_Chouhan,Lami_muPCD_Labram,Lami_muPCD_Guse,Lami_muPCD_Bi} and low dimensional semiconductor materials including carbon nanotubes \cite{FreitagNL2003,LuNT2006}, graphene \cite{VaskoPRB2008,DochertyNatCom2012}, transition metal dichalcogenides \cite{XiaNatPhot2014}, and black phosphorus \cite{LiuAdvMat2016,MiaoACSNano2017}. For such materials sensitivity to material parameters, as well as a sensitive (i.e. background reflection free) measurement of the $\mu$-PCD signal are important rather than the large throughput study of an industrial investigation. 

\section{Summary}
In summary, we presented an improved approach to detect photoinduced conductivity in semiconductors using microwave resonators. Previous studies with microwave resonators have yielded material parameters after involved modeling or with a slow time dynamics (beyond a few ms-second). Our approach yields directly the resonator parameters, which are in turn related to the material parameters. It is based on the detection of the transient response of a microwave cavity. While the method encompasses all the known benefits of resonators in terms of sensitivity and accuracy, its ultimate time resolution is the resonator time constant which can be as low as a few ns.

\section*{Acknowledgements}
The Authors are indebted to Dr. Dario Quintavalle from the Semilab Semiconductor Physics Laboratory Ltd. for useful advises and for providing manufacturing grade silicon wafer samples. This work was supported by the Hungarian National Research, Development and Innovation Office (NKFIH) Grant Nrs. 2017-1.2.1-NKP-2017-00001 and K119442, and by the BME-Nanonotechnology FIKP grant of EMMI (BME FIKP-NAT).


\begin{thebibliography}{56}
\expandafter\ifx\csname natexlab\endcsname\relax\def\natexlab#1{#1}\fi
\expandafter\ifx\csname bibnamefont\endcsname\relax
  \def\bibnamefont#1{#1}\fi
\expandafter\ifx\csname bibfnamefont\endcsname\relax
  \def\bibfnamefont#1{#1}\fi
\expandafter\ifx\csname citenamefont\endcsname\relax
  \def\citenamefont#1{#1}\fi
\expandafter\ifx\csname url\endcsname\relax
  \def\url#1{\texttt{#1}}\fi
\expandafter\ifx\csname urlprefix\endcsname\relax\def\urlprefix{URL }\fi
\providecommand{\bibinfo}[2]{#2}
\providecommand{\eprint}[2][]{\url{#2}}

\bibitem[{\citenamefont{Ohsawa et~al.}(1983)\citenamefont{Ohsawa, Honda,
  Takizawa, and Toyokura}}]{Ohsawa1983}
\bibinfo{author}{\bibfnamefont{A.}~\bibnamefont{Ohsawa}},
  \bibinfo{author}{\bibfnamefont{K.}~\bibnamefont{Honda}},
  \bibinfo{author}{\bibfnamefont{R.}~\bibnamefont{Takizawa}}, \bibnamefont{and}
  \bibinfo{author}{\bibfnamefont{N.}~\bibnamefont{Toyokura}},
  \bibinfo{journal}{Review of Scientific Instruments}
  \textbf{\bibinfo{volume}{54}}, \bibinfo{pages}{210 } (\bibinfo{year}{1983}).

\bibitem[{\citenamefont{Kunst and Beck}(1986)}]{Kunst1986}
\bibinfo{author}{\bibfnamefont{M.}~\bibnamefont{Kunst}} \bibnamefont{and}
  \bibinfo{author}{\bibfnamefont{G.}~\bibnamefont{Beck}},
  \bibinfo{journal}{Journal of Applied Physics} \textbf{\bibinfo{volume}{60}},
  \bibinfo{pages}{3558} (\bibinfo{year}{1986}).

\bibitem[{\citenamefont{Lauer et~al.}(2008)\citenamefont{Lauer, Laades,
  Ubensee, Metzner, and Lawerenz}}]{muPCD_summary1}
\bibinfo{author}{\bibfnamefont{K.}~\bibnamefont{Lauer}},
  \bibinfo{author}{\bibfnamefont{A.}~\bibnamefont{Laades}},
  \bibinfo{author}{\bibfnamefont{H.}~\bibnamefont{Ubensee}},
  \bibinfo{author}{\bibfnamefont{H.}~\bibnamefont{Metzner}}, \bibnamefont{and}
  \bibinfo{author}{\bibfnamefont{A.}~\bibnamefont{Lawerenz}},
  \bibinfo{journal}{Journal of Applied Physics} \textbf{\bibinfo{volume}{104}},
  \bibinfo{pages}{104503 } (\bibinfo{year}{2008}).

\bibitem[{\citenamefont{Berger et~al.}(2011)\citenamefont{Berger, Schüler,
  Anger, Gründig-Wendrock, Niklas, and Dornich}}]{muPCD_Co_characterization}
\bibinfo{author}{\bibfnamefont{B.}~\bibnamefont{Berger}},
  \bibinfo{author}{\bibfnamefont{N.}~\bibnamefont{Schüler}},
  \bibinfo{author}{\bibfnamefont{S.}~\bibnamefont{Anger}},
  \bibinfo{author}{\bibfnamefont{B.}~\bibnamefont{Gründig-Wendrock}},
  \bibinfo{author}{\bibfnamefont{J.~R.} \bibnamefont{Niklas}},
  \bibnamefont{and} \bibinfo{author}{\bibfnamefont{K.}~\bibnamefont{Dornich}},
  \bibinfo{journal}{physica status solidi (a)} \textbf{\bibinfo{volume}{208}},
  \bibinfo{pages}{769} (\bibinfo{year}{2011}).

\bibitem[{\citenamefont{Chouhan et~al.}(2017)\citenamefont{Chouhan, Jasti,
  Hadke, Raghavan, and Avasthi}}]{Lami_muPCD_Chouhan}
\bibinfo{author}{\bibfnamefont{A.~S.} \bibnamefont{Chouhan}},
  \bibinfo{author}{\bibfnamefont{N.~P.} \bibnamefont{Jasti}},
  \bibinfo{author}{\bibfnamefont{S.}~\bibnamefont{Hadke}},
  \bibinfo{author}{\bibfnamefont{S.}~\bibnamefont{Raghavan}}, \bibnamefont{and}
  \bibinfo{author}{\bibfnamefont{S.}~\bibnamefont{Avasthi}},
  \bibinfo{journal}{Current Applied Physics} \textbf{\bibinfo{volume}{17}},
  \bibinfo{pages}{1335 } (\bibinfo{year}{2017}).

\bibitem[{\citenamefont{Labram and Chabinyc}(2017)}]{Lami_muPCD_Labram}
\bibinfo{author}{\bibfnamefont{J.~G.} \bibnamefont{Labram}} \bibnamefont{and}
  \bibinfo{author}{\bibfnamefont{M.~L.} \bibnamefont{Chabinyc}},
  \bibinfo{journal}{Journal of Applied Physics} \textbf{\bibinfo{volume}{122}},
  \bibinfo{pages}{065501} (\bibinfo{year}{2017}).

\bibitem[{\citenamefont{Guse et~al.}(2016)\citenamefont{Guse, Soufiani, Jiang,
  Kim, Cheng, Schmidt, Ho-Baillie, and McCamey}}]{Lami_muPCD_Guse}
\bibinfo{author}{\bibfnamefont{J.~A.} \bibnamefont{Guse}},
  \bibinfo{author}{\bibfnamefont{A.~M.} \bibnamefont{Soufiani}},
  \bibinfo{author}{\bibfnamefont{L.}~\bibnamefont{Jiang}},
  \bibinfo{author}{\bibfnamefont{J.}~\bibnamefont{Kim}},
  \bibinfo{author}{\bibfnamefont{Y.-B.} \bibnamefont{Cheng}},
  \bibinfo{author}{\bibfnamefont{T.~W.} \bibnamefont{Schmidt}},
  \bibinfo{author}{\bibfnamefont{A.}~\bibnamefont{Ho-Baillie}},
  \bibnamefont{and} \bibinfo{author}{\bibfnamefont{D.~R.}
  \bibnamefont{McCamey}}, \bibinfo{journal}{Phys. Chem. Chem. Phys.}
  \textbf{\bibinfo{volume}{18}}, \bibinfo{pages}{12043} (\bibinfo{year}{2016}).

\bibitem[{\citenamefont{Bi et~al.}(2016)\citenamefont{Bi, Hutter, Fang, Dong,
  Huang, and Savenije}}]{Lami_muPCD_Bi}
\bibinfo{author}{\bibfnamefont{Y.}~\bibnamefont{Bi}},
  \bibinfo{author}{\bibfnamefont{E.~M.} \bibnamefont{Hutter}},
  \bibinfo{author}{\bibfnamefont{Y.}~\bibnamefont{Fang}},
  \bibinfo{author}{\bibfnamefont{Q.}~\bibnamefont{Dong}},
  \bibinfo{author}{\bibfnamefont{J.}~\bibnamefont{Huang}}, \bibnamefont{and}
  \bibinfo{author}{\bibfnamefont{T.~J.} \bibnamefont{Savenije}},
  \bibinfo{journal}{The Journal of Physical Chemistry Letters}
  \textbf{\bibinfo{volume}{7}}, \bibinfo{pages}{923} (\bibinfo{year}{2016}).

\bibitem[{\citenamefont{Novikov et~al.}(2010)\citenamefont{Novikov, A.~Marinin,
  and V.~Rabenok}}]{Novikov_CdTe}
\bibinfo{author}{\bibfnamefont{G.}~\bibnamefont{Novikov}},
  \bibinfo{author}{\bibfnamefont{A.}~\bibnamefont{A.~Marinin}},
  \bibnamefont{and}
  \bibinfo{author}{\bibfnamefont{E.}~\bibnamefont{V.~Rabenok}},
  \bibinfo{journal}{Instruments and Experimental Techniques}
  \textbf{\bibinfo{volume}{53}}, \bibinfo{pages}{233} (\bibinfo{year}{2010}).

\bibitem[{\citenamefont{Freitag et~al.}(2003)\citenamefont{Freitag, Martin,
  Misewich, Martel, and Avouris}}]{FreitagNL2003}
\bibinfo{author}{\bibfnamefont{M.}~\bibnamefont{Freitag}},
  \bibinfo{author}{\bibfnamefont{Y.}~\bibnamefont{Martin}},
  \bibinfo{author}{\bibfnamefont{J.~A.} \bibnamefont{Misewich}},
  \bibinfo{author}{\bibfnamefont{R.}~\bibnamefont{Martel}}, \bibnamefont{and}
  \bibinfo{author}{\bibfnamefont{P.}~\bibnamefont{Avouris}},
  \bibinfo{journal}{Nano Lett.} \textbf{\bibinfo{volume}{3}},
  \bibinfo{pages}{1067} (\bibinfo{year}{2003}).

\bibitem[{\citenamefont{Lu and Panchapakesan}(2006)}]{LuNT2006}
\bibinfo{author}{\bibfnamefont{S.}~\bibnamefont{Lu}} \bibnamefont{and}
  \bibinfo{author}{\bibfnamefont{B.}~\bibnamefont{Panchapakesan}},
  \bibinfo{journal}{Nanotechnology} \textbf{\bibinfo{volume}{17}},
  \bibinfo{pages}{1843} (\bibinfo{year}{2006}).

\bibitem[{\citenamefont{Vasko and Ryzhii}(2008)}]{VaskoPRB2008}
\bibinfo{author}{\bibfnamefont{F.~T.} \bibnamefont{Vasko}} \bibnamefont{and}
  \bibinfo{author}{\bibfnamefont{V.}~\bibnamefont{Ryzhii}},
  \bibinfo{journal}{Phys. Rev. B} \textbf{\bibinfo{volume}{77}},
  \bibinfo{pages}{195433} (\bibinfo{year}{2008}).

\bibitem[{\citenamefont{Docherty et~al.}(2012)\citenamefont{Docherty, Lin,
  Joyce, Nicholas, Herz, Li, and Johnston}}]{DochertyNatCom2012}
\bibinfo{author}{\bibfnamefont{C.~J.} \bibnamefont{Docherty}},
  \bibinfo{author}{\bibfnamefont{C.-T.} \bibnamefont{Lin}},
  \bibinfo{author}{\bibfnamefont{H.~J.} \bibnamefont{Joyce}},
  \bibinfo{author}{\bibfnamefont{R.~J.} \bibnamefont{Nicholas}},
  \bibinfo{author}{\bibfnamefont{L.~M.} \bibnamefont{Herz}},
  \bibinfo{author}{\bibfnamefont{L.-J.} \bibnamefont{Li}}, \bibnamefont{and}
  \bibinfo{author}{\bibfnamefont{M.~B.} \bibnamefont{Johnston}},
  \bibinfo{journal}{Nature Communications} \textbf{\bibinfo{volume}{3}},
  \bibinfo{pages}{1228} (\bibinfo{year}{2012}).

\bibitem[{\citenamefont{Xia et~al.}(2014)\citenamefont{Xia, Wang, Xiao, Dubey,
  and Ramasubramaniam}}]{XiaNatPhot2014}
\bibinfo{author}{\bibfnamefont{F.}~\bibnamefont{Xia}},
  \bibinfo{author}{\bibfnamefont{H.}~\bibnamefont{Wang}},
  \bibinfo{author}{\bibfnamefont{D.}~\bibnamefont{Xiao}},
  \bibinfo{author}{\bibfnamefont{M.}~\bibnamefont{Dubey}}, \bibnamefont{and}
  \bibinfo{author}{\bibfnamefont{A.}~\bibnamefont{Ramasubramaniam}},
  \bibinfo{journal}{Nature Photonics} \textbf{\bibinfo{volume}{8}},
  \bibinfo{pages}{899} (\bibinfo{year}{2014}).

\bibitem[{\citenamefont{Liu et~al.}(2016)\citenamefont{Liu, Zhu, You, Liang,
  Zheng, Zhou, Fu, He, Zeng, Fan et~al.}}]{LiuAdvMat2016}
\bibinfo{author}{\bibfnamefont{F.}~\bibnamefont{Liu}},
  \bibinfo{author}{\bibfnamefont{C.}~\bibnamefont{Zhu}},
  \bibinfo{author}{\bibfnamefont{L.}~\bibnamefont{You}},
  \bibinfo{author}{\bibfnamefont{S.-J.} \bibnamefont{Liang}},
  \bibinfo{author}{\bibfnamefont{S.}~\bibnamefont{Zheng}},
  \bibinfo{author}{\bibfnamefont{J.}~\bibnamefont{Zhou}},
  \bibinfo{author}{\bibfnamefont{Q.}~\bibnamefont{Fu}},
  \bibinfo{author}{\bibfnamefont{Y.}~\bibnamefont{He}},
  \bibinfo{author}{\bibfnamefont{Q.}~\bibnamefont{Zeng}},
  \bibinfo{author}{\bibfnamefont{H.~J.} \bibnamefont{Fan}},
  \bibnamefont{et~al.}, \bibinfo{journal}{Advanced Materials}
  \textbf{\bibinfo{volume}{28}}, \bibinfo{pages}{7768} (\bibinfo{year}{2016}).

\bibitem[{\citenamefont{Miao et~al.}(2017)\citenamefont{Miao, Song, Li, Cai,
  Zhang, Hu, Dong, and Wang}}]{MiaoACSNano2017}
\bibinfo{author}{\bibfnamefont{J.}~\bibnamefont{Miao}},
  \bibinfo{author}{\bibfnamefont{B.}~\bibnamefont{Song}},
  \bibinfo{author}{\bibfnamefont{Q.}~\bibnamefont{Li}},
  \bibinfo{author}{\bibfnamefont{L.}~\bibnamefont{Cai}},
  \bibinfo{author}{\bibfnamefont{S.}~\bibnamefont{Zhang}},
  \bibinfo{author}{\bibfnamefont{W.}~\bibnamefont{Hu}},
  \bibinfo{author}{\bibfnamefont{L.}~\bibnamefont{Dong}}, \bibnamefont{and}
  \bibinfo{author}{\bibfnamefont{C.}~\bibnamefont{Wang}}, \bibinfo{journal}{ACS
  Nano} \textbf{\bibinfo{volume}{11}}, \bibinfo{pages}{6048}
  (\bibinfo{year}{2017}).

\bibitem[{\citenamefont{Infelta et~al.}(1977)\citenamefont{Infelta, de~Haas,
  and Warman}}]{Warman_RadPhysChem_1977}
\bibinfo{author}{\bibfnamefont{P.~P.} \bibnamefont{Infelta}},
  \bibinfo{author}{\bibfnamefont{M.~P.} \bibnamefont{de~Haas}},
  \bibnamefont{and} \bibinfo{author}{\bibfnamefont{J.~M.}
  \bibnamefont{Warman}}, \bibinfo{journal}{Radiation Physics and Chemistry
  (1977)} \textbf{\bibinfo{volume}{10}}, \bibinfo{pages}{353 }
  (\bibinfo{year}{1977}), ISSN \bibinfo{issn}{0146-5724}.

\bibitem[{\citenamefont{Fessenden et~al.}(1982)\citenamefont{Fessenden, Carton,
  Shimamori, and Scaiano}}]{fessenden1982_AFC_muPCD_meas}
\bibinfo{author}{\bibfnamefont{R.~W.} \bibnamefont{Fessenden}},
  \bibinfo{author}{\bibfnamefont{P.~M.} \bibnamefont{Carton}},
  \bibinfo{author}{\bibfnamefont{H.}~\bibnamefont{Shimamori}},
  \bibnamefont{and} \bibinfo{author}{\bibfnamefont{J.~C.}
  \bibnamefont{Scaiano}}, \bibinfo{journal}{The Journal of Physical Chemistry}
  \textbf{\bibinfo{volume}{86}}, \bibinfo{pages}{3803} (\bibinfo{year}{1982}).

\bibitem[{\citenamefont{Reid et~al.}(2017)\citenamefont{Reid, Moore, Li, Zhao,
  Yan, Zhu, and Rumbles}}]{Reid_2017}
\bibinfo{author}{\bibfnamefont{O.~G.} \bibnamefont{Reid}},
  \bibinfo{author}{\bibfnamefont{D.~T.} \bibnamefont{Moore}},
  \bibinfo{author}{\bibfnamefont{Z.}~\bibnamefont{Li}},
  \bibinfo{author}{\bibfnamefont{D.}~\bibnamefont{Zhao}},
  \bibinfo{author}{\bibfnamefont{Y.}~\bibnamefont{Yan}},
  \bibinfo{author}{\bibfnamefont{K.}~\bibnamefont{Zhu}}, \bibnamefont{and}
  \bibinfo{author}{\bibfnamefont{G.}~\bibnamefont{Rumbles}},
  \bibinfo{journal}{Journal of Physics D: Applied Physics}
  \textbf{\bibinfo{volume}{50}}, \bibinfo{pages}{493002}
  (\bibinfo{year}{2017}).

\bibitem[{\citenamefont{Pozar}(2004)}]{PozarBook}
\bibinfo{author}{\bibfnamefont{D.~M.} \bibnamefont{Pozar}},
  \emph{\bibinfo{title}{Microwave Engineering}} (\bibinfo{publisher}{John Wiley
  \& Sons, Inc.}, \bibinfo{year}{2004}).

\bibitem[{\citenamefont{{Cetinoneri} et~al.}(2010)\citenamefont{{Cetinoneri},
  {Atesal}, {Kroeger}, {Tepper}, {Losee}, {Hicks}, {Rasmussen}, and
  {Rebeiz}}}]{cetinoneri2010_time_resolved_muPCD}
\bibinfo{author}{\bibfnamefont{B.}~\bibnamefont{{Cetinoneri}}},
  \bibinfo{author}{\bibfnamefont{Y.~A.} \bibnamefont{{Atesal}}},
  \bibinfo{author}{\bibfnamefont{R.~A.} \bibnamefont{{Kroeger}}},
  \bibinfo{author}{\bibfnamefont{G.}~\bibnamefont{{Tepper}}},
  \bibinfo{author}{\bibfnamefont{J.}~\bibnamefont{{Losee}}},
  \bibinfo{author}{\bibfnamefont{C.}~\bibnamefont{{Hicks}}},
  \bibinfo{author}{\bibfnamefont{M.}~\bibnamefont{{Rasmussen}}},
  \bibnamefont{and} \bibinfo{author}{\bibfnamefont{G.~M.}
  \bibnamefont{{Rebeiz}}}, in \emph{\bibinfo{booktitle}{2010 IEEE MTT-S
  International Microwave Symposium}} (\bibinfo{year}{2010}), pp.
  \bibinfo{pages}{469--472}, ISSN \bibinfo{issn}{0149-645X}.

\bibitem[{\citenamefont{Subramanian et~al.}(1998)\citenamefont{Subramanian,
  Murthy, and Sobhanadri}}]{SubramanianJAP1998}
\bibinfo{author}{\bibfnamefont{V.}~\bibnamefont{Subramanian}},
  \bibinfo{author}{\bibfnamefont{V.~R.~K.} \bibnamefont{Murthy}},
  \bibnamefont{and}
  \bibinfo{author}{\bibfnamefont{J.}~\bibnamefont{Sobhanadri}},
  \bibinfo{journal}{Journal of Applied Physics} \textbf{\bibinfo{volume}{83}},
  \bibinfo{pages}{837} (\bibinfo{year}{1998}).

\bibitem[{\citenamefont{Amato}(1982)}]{Amato}
\bibinfo{author}{\bibfnamefont{J.}~\bibnamefont{Amato}},
  \bibinfo{journal}{Review of Scientific Instruments}
  \textbf{\bibinfo{volume}{53}}, \bibinfo{pages}{776} (\bibinfo{year}{1982}).

\bibitem[{\citenamefont{Eckstrom et~al.}(1987)\citenamefont{Eckstrom, Williams,
  and Dickinson}}]{Eckstrom}
\bibinfo{author}{\bibfnamefont{D.~J.} \bibnamefont{Eckstrom}},
  \bibinfo{author}{\bibfnamefont{M.~S.} \bibnamefont{Williams}},
  \bibnamefont{and}
  \bibinfo{author}{\bibfnamefont{J.}~\bibnamefont{Dickinson}},
  \bibinfo{journal}{Review of Scientific Instruments}
  \textbf{\bibinfo{volume}{58}}, \bibinfo{pages}{2244} (\bibinfo{year}{1987}).

\bibitem[{\citenamefont{Kessick et~al.}(2000)\citenamefont{Kessick, Tepper,
  Lee, and James}}]{Tepper_detector_2000}
\bibinfo{author}{\bibfnamefont{R.}~\bibnamefont{Kessick}},
  \bibinfo{author}{\bibfnamefont{G.}~\bibnamefont{Tepper}},
  \bibinfo{author}{\bibfnamefont{E.}~\bibnamefont{Lee}}, \bibnamefont{and}
  \bibinfo{author}{\bibfnamefont{R.}~\bibnamefont{James}},
  \bibinfo{journal}{Journal of Applied Physics} \textbf{\bibinfo{volume}{87}},
  \bibinfo{pages}{2408} (\bibinfo{year}{2000}).

\bibitem[{\citenamefont{Tepper and Losee}(2001)}]{Tepper_detector_2001}
\bibinfo{author}{\bibfnamefont{G.}~\bibnamefont{Tepper}} \bibnamefont{and}
  \bibinfo{author}{\bibfnamefont{J.}~\bibnamefont{Losee}},
  \bibinfo{journal}{Nuclear Instruments and Methods in Physics Research Section
  A: Accelerators, Spectrometers, Detectors and Associated Equipment}
  \textbf{\bibinfo{volume}{458}}, \bibinfo{pages}{472 } (\bibinfo{year}{2001}),
  ISSN \bibinfo{issn}{0168-9002}, \bibinfo{note}{proc. 11th Inbt. Workshop on
  Room Temperature Semiconductor X- and Gamma-Ray Detectors and Associated
  Electronics}.

\bibitem[{\citenamefont{Braggio et~al.}(2014)\citenamefont{Braggio, Carugno,
  Lombardi, Ruoso, and Sirugudu}}]{Braggio_detector_2014}
\bibinfo{author}{\bibfnamefont{C.}~\bibnamefont{Braggio}},
  \bibinfo{author}{\bibfnamefont{G.}~\bibnamefont{Carugno}},
  \bibinfo{author}{\bibfnamefont{A.}~\bibnamefont{Lombardi}},
  \bibinfo{author}{\bibfnamefont{G.}~\bibnamefont{Ruoso}}, \bibnamefont{and}
  \bibinfo{author}{\bibfnamefont{R.}~\bibnamefont{Sirugudu}},
  \bibinfo{journal}{Journal of Applied Physics} \textbf{\bibinfo{volume}{116}},
  \bibinfo{pages}{044513} (\bibinfo{year}{2014}).

\bibitem[{\citenamefont{Petersan and Anlage}(1998)}]{PetersanAnlage}
\bibinfo{author}{\bibfnamefont{P.~J.} \bibnamefont{Petersan}} \bibnamefont{and}
  \bibinfo{author}{\bibfnamefont{S.~M.} \bibnamefont{Anlage}},
  \bibinfo{journal}{Journal of Applied Physics} \textbf{\bibinfo{volume}{84}},
  \bibinfo{pages}{3392} (\bibinfo{year}{1998}).

\bibitem[{\citenamefont{Luiten}(2005)}]{LuitenReview}
\bibinfo{author}{\bibfnamefont{A.}~\bibnamefont{Luiten}},
  \emph{\bibinfo{title}{Q-Factor Measurements}} (\bibinfo{publisher}{John Wiley
  \& Sons, Inc.}, \bibinfo{year}{2005}), Encyclopedia of RF and Microwave
  Engineering.

\bibitem[{\citenamefont{Kajfez}(2005)}]{KajfezReview}
\bibinfo{author}{\bibfnamefont{D.}~\bibnamefont{Kajfez}},
  \emph{\bibinfo{title}{Q-Factor}} (\bibinfo{publisher}{John Wiley \& Sons,
  Inc.}, \bibinfo{year}{2005}), Encyclopedia of RF and Microwave Engineering.

\bibitem[{\citenamefont{Gy\"{u}re et~al.}(2015)\citenamefont{Gy\"{u}re,
  G.~M\'{a}rkus, Bern\'{a}th, Mur\'{a}nyi, and Simon}}]{GyureRSI}
\bibinfo{author}{\bibfnamefont{B.}~\bibnamefont{Gy\"{u}re}},
  \bibinfo{author}{\bibfnamefont{B.}~\bibnamefont{G.~M\'{a}rkus}},
  \bibinfo{author}{\bibfnamefont{B.}~\bibnamefont{Bern\'{a}th}},
  \bibinfo{author}{\bibfnamefont{F.}~\bibnamefont{Mur\'{a}nyi}},
  \bibnamefont{and} \bibinfo{author}{\bibfnamefont{F.}~\bibnamefont{Simon}},
  \bibinfo{journal}{Review of Scientific Instruments}
  \textbf{\bibinfo{volume}{86}}, \bibinfo{pages}{094702}
  (\bibinfo{year}{2015}).

\bibitem[{\citenamefont{Gyüre-Garami et~al.}(2018)\citenamefont{Gyüre-Garami,
  Sági, Márkus, and Simon}}]{GyureGaramiRSI}
\bibinfo{author}{\bibfnamefont{B.}~\bibnamefont{Gyüre-Garami}},
  \bibinfo{author}{\bibfnamefont{O.}~\bibnamefont{Sági}},
  \bibinfo{author}{\bibfnamefont{B.~G.} \bibnamefont{Márkus}},
  \bibnamefont{and} \bibinfo{author}{\bibfnamefont{F.}~\bibnamefont{Simon}},
  \bibinfo{journal}{Review of Scientific Instruments}
  \textbf{\bibinfo{volume}{89}}, \bibinfo{pages}{113903}
  (\bibinfo{year}{2018}).

\bibitem[{\citenamefont{Ernst}(1992)}]{Ernst}
\bibinfo{author}{\bibfnamefont{R.~R.} \bibnamefont{Ernst}},
  \bibinfo{journal}{Angewandte Chemie-International Edition in English}
  \textbf{\bibinfo{volume}{31}}, \bibinfo{pages}{805} (\bibinfo{year}{1992}).

\bibitem[{\citenamefont{Hodges et~al.}(2004)\citenamefont{Hodges, Layer,
  Miller, and Scace}}]{optics1}
\bibinfo{author}{\bibfnamefont{J.}~\bibnamefont{Hodges}},
  \bibinfo{author}{\bibfnamefont{H.}~\bibnamefont{Layer}},
  \bibinfo{author}{\bibfnamefont{W.}~\bibnamefont{Miller}}, \bibnamefont{and}
  \bibinfo{author}{\bibfnamefont{G.}~\bibnamefont{Scace}},
  \bibinfo{journal}{Review of Scientific Instruments}
  \textbf{\bibinfo{volume}{75}}, \bibinfo{pages}{849} (\bibinfo{year}{2004}).

\bibitem[{\citenamefont{Hodges and Ciurylo}(2005)}]{optics2}
\bibinfo{author}{\bibfnamefont{J.}~\bibnamefont{Hodges}} \bibnamefont{and}
  \bibinfo{author}{\bibfnamefont{R.}~\bibnamefont{Ciurylo}},
  \bibinfo{journal}{Review of Scientific Instruments}
  \textbf{\bibinfo{volume}{76}}, \bibinfo{pages}{023112}
  (\bibinfo{year}{2005}).

\bibitem[{\citenamefont{Cygan et~al.}(2011)\citenamefont{Cygan, Lisak,
  Maslowski, Bielska, Wojtewicz, Domyslawska, Trawinski, Ciurylo, Abe, and
  Hodges}}]{optics3}
\bibinfo{author}{\bibfnamefont{A.}~\bibnamefont{Cygan}},
  \bibinfo{author}{\bibfnamefont{D.}~\bibnamefont{Lisak}},
  \bibinfo{author}{\bibfnamefont{P.}~\bibnamefont{Maslowski}},
  \bibinfo{author}{\bibfnamefont{K.}~\bibnamefont{Bielska}},
  \bibinfo{author}{\bibfnamefont{S.}~\bibnamefont{Wojtewicz}},
  \bibinfo{author}{\bibfnamefont{J.}~\bibnamefont{Domyslawska}},
  \bibinfo{author}{\bibfnamefont{R.~S.} \bibnamefont{Trawinski}},
  \bibinfo{author}{\bibfnamefont{R.}~\bibnamefont{Ciurylo}},
  \bibinfo{author}{\bibfnamefont{H.}~\bibnamefont{Abe}}, \bibnamefont{and}
  \bibinfo{author}{\bibfnamefont{J.~T.} \bibnamefont{Hodges}},
  \bibinfo{journal}{Review of Scientific Instruments}
  \textbf{\bibinfo{volume}{82}}, \bibinfo{pages}{063107}
  (\bibinfo{year}{2011}).

\bibitem[{\citenamefont{Truong et~al.}(2013)\citenamefont{Truong, Long, Cygan,
  Lisak, van Zee, and Hodges}}]{optics4}
\bibinfo{author}{\bibfnamefont{G.~W.} \bibnamefont{Truong}},
  \bibinfo{author}{\bibfnamefont{D.~A.} \bibnamefont{Long}},
  \bibinfo{author}{\bibfnamefont{A.}~\bibnamefont{Cygan}},
  \bibinfo{author}{\bibfnamefont{D.}~\bibnamefont{Lisak}},
  \bibinfo{author}{\bibfnamefont{R.~D.} \bibnamefont{van Zee}},
  \bibnamefont{and} \bibinfo{author}{\bibfnamefont{J.~T.}
  \bibnamefont{Hodges}}, \bibinfo{journal}{J. Chem. Phys.}
  \textbf{\bibinfo{volume}{138}}, \bibinfo{pages}{094201}
  (\bibinfo{year}{2013}).

\bibitem[{\citenamefont{Schmitt and Zimmer}(1966)}]{SchmittZimmer}
\bibinfo{author}{\bibfnamefont{H.~J.} \bibnamefont{Schmitt}} \bibnamefont{and}
  \bibinfo{author}{\bibfnamefont{H.}~\bibnamefont{Zimmer}},
  \bibinfo{journal}{IEEE Transactions on Microwave Theory and Techniques}
  \textbf{\bibinfo{volume}{MT14}}, \bibinfo{pages}{206} (\bibinfo{year}{1966}).

\bibitem[{\citenamefont{Gallagher}(1979)}]{Gallagher}
\bibinfo{author}{\bibfnamefont{W.}~\bibnamefont{Gallagher}},
  \bibinfo{journal}{IEEE Transactions on Nuclear Science}
  \textbf{\bibinfo{volume}{26}}, \bibinfo{pages}{4277} (\bibinfo{year}{1979}).

\bibitem[{\citenamefont{Komachi and Tanaka}(1974)}]{KomachiTanaka}
\bibinfo{author}{\bibfnamefont{Y.}~\bibnamefont{Komachi}} \bibnamefont{and}
  \bibinfo{author}{\bibfnamefont{S.}~\bibnamefont{Tanaka}},
  \bibinfo{journal}{Journal of Physics E: Scientific Instruments}
  \textbf{\bibinfo{volume}{7}}, \bibinfo{pages}{905} (\bibinfo{year}{1974}).

\bibitem[{\citenamefont{Quine et~al.}(2011)\citenamefont{Quine, Mitchell, and
  Eaton}}]{EatonTransient}
\bibinfo{author}{\bibfnamefont{R.~W.} \bibnamefont{Quine}},
  \bibinfo{author}{\bibfnamefont{D.~G.} \bibnamefont{Mitchell}},
  \bibnamefont{and} \bibinfo{author}{\bibfnamefont{G.~R.} \bibnamefont{Eaton}},
  \bibinfo{journal}{Concepts in Magnetic Resonance Part B: Magnetic Resonance
  Engineering} \textbf{\bibinfo{volume}{39B}}, \bibinfo{pages}{43}
  (\bibinfo{year}{2011}).

\bibitem[{\citenamefont{Connes and Connes}(1966)}]{Connes}
\bibinfo{author}{\bibfnamefont{J.}~\bibnamefont{Connes}} \bibnamefont{and}
  \bibinfo{author}{\bibfnamefont{P.}~\bibnamefont{Connes}},
  \bibinfo{journal}{J. Opt. Soc. Am.} \textbf{\bibinfo{volume}{56}}
  (\bibinfo{year}{1966}).

\bibitem[{\citenamefont{Fellgett}(1949)}]{Fellgett}
\bibinfo{author}{\bibfnamefont{P.~B.} \bibnamefont{Fellgett}},
  \emph{\bibinfo{title}{{T}heory of {I}nfra-{R}ed {S}ensitivities and its
  {A}pplication to {I}nvestigations of {S}tellar {R}adiation in the {N}ear
  {I}nfra-{R}ed ({P}h{D} thesis).}} (\bibinfo{publisher}{University of
  Cambridge}, \bibinfo{year}{1949}).

\bibitem[{\citenamefont{G.~Márkus et~al.}(2018)\citenamefont{G.~Márkus,
  Gyüre-Garami, Sági, Csősz, Karsa, Márkus, and Simon}}]{MarkusGyurePSSB}
\bibinfo{author}{\bibfnamefont{B.}~\bibnamefont{G.~Márkus}},
  \bibinfo{author}{\bibfnamefont{B.}~\bibnamefont{Gyüre-Garami}},
  \bibinfo{author}{\bibfnamefont{O.}~\bibnamefont{Sági}},
  \bibinfo{author}{\bibfnamefont{G.}~\bibnamefont{Csősz}},
  \bibinfo{author}{\bibfnamefont{A.}~\bibnamefont{Karsa}},
  \bibinfo{author}{\bibfnamefont{F.}~\bibnamefont{Márkus}}, \bibnamefont{and}
  \bibinfo{author}{\bibfnamefont{F.}~\bibnamefont{Simon}},
  \bibinfo{journal}{physica status solidi (b)} \textbf{\bibinfo{volume}{255}}
  (\bibinfo{year}{2018}).

\bibitem[{\citenamefont{Gresits et~al.}(2019)\citenamefont{Gresits,
  Thur{\'{o}}czy, S{\'{a}}gi, Homolya, Bagam{\'{e}}ry, Gaj{\'{a}}ri, Babos,
  Major, M{\'{a}}rkus, and Simon}}]{GresitsJPD}
\bibinfo{author}{\bibfnamefont{I.}~\bibnamefont{Gresits}},
  \bibinfo{author}{\bibfnamefont{G.}~\bibnamefont{Thur{\'{o}}czy}},
  \bibinfo{author}{\bibfnamefont{O.}~\bibnamefont{S{\'{a}}gi}},
  \bibinfo{author}{\bibfnamefont{I.}~\bibnamefont{Homolya}},
  \bibinfo{author}{\bibfnamefont{G.}~\bibnamefont{Bagam{\'{e}}ry}},
  \bibinfo{author}{\bibfnamefont{D.}~\bibnamefont{Gaj{\'{a}}ri}},
  \bibinfo{author}{\bibfnamefont{M.}~\bibnamefont{Babos}},
  \bibinfo{author}{\bibfnamefont{P.}~\bibnamefont{Major}},
  \bibinfo{author}{\bibfnamefont{B.~G.} \bibnamefont{M{\'{a}}rkus}},
  \bibnamefont{and} \bibinfo{author}{\bibfnamefont{F.}~\bibnamefont{Simon}},
  \bibinfo{journal}{Journal of Physics D: Applied Physics}
  \textbf{\bibinfo{volume}{52}}, \bibinfo{pages}{375401}
  (\bibinfo{year}{2019}).

\bibitem[{\citenamefont{Chen et~al.}(2004)\citenamefont{Chen, Ong, Neo,
  Varadan, and Varadan}}]{chen2004microwave}
\bibinfo{author}{\bibfnamefont{L.}~\bibnamefont{Chen}},
  \bibinfo{author}{\bibfnamefont{C.}~\bibnamefont{Ong}},
  \bibinfo{author}{\bibfnamefont{C.}~\bibnamefont{Neo}},
  \bibinfo{author}{\bibfnamefont{V.}~\bibnamefont{Varadan}}, \bibnamefont{and}
  \bibinfo{author}{\bibfnamefont{V.}~\bibnamefont{Varadan}},
  \emph{\bibinfo{title}{Microwave Electronics: Measurement and Materials
  Characterization}} (\bibinfo{publisher}{John Wiley \& Sons, Inc.},
  \bibinfo{year}{2004}).

\bibitem[{\citenamefont{Ristic et~al.}(2004)\citenamefont{Ristic, Prijic, and
  Prijic}}]{EpsilonVariation}
\bibinfo{author}{\bibfnamefont{S.}~\bibnamefont{Ristic}},
  \bibinfo{author}{\bibfnamefont{A.}~\bibnamefont{Prijic}}, \bibnamefont{and}
  \bibinfo{author}{\bibfnamefont{Z.}~\bibnamefont{Prijic}},
  \bibinfo{journal}{Serbian Journal of Electrical Engineering}
  \textbf{\bibinfo{volume}{1}}, \bibinfo{pages}{237} (\bibinfo{year}{2004}).

\bibitem[{\citenamefont{Buravov and Shchegolev}(1971)}]{buravov71}
\bibinfo{author}{\bibfnamefont{L.~I.} \bibnamefont{Buravov}} \bibnamefont{and}
  \bibinfo{author}{\bibfnamefont{I.~F.} \bibnamefont{Shchegolev}},
  \bibinfo{journal}{Instrum.\ Exp.\ Tech.} \textbf{\bibinfo{volume}{14}},
  \bibinfo{pages}{528} (\bibinfo{year}{1971}).

\bibitem[{\citenamefont{Klein et~al.}(1993)\citenamefont{Klein, Donovan,
  Dressel, and Gr\"uner}}]{Gruner1}
\bibinfo{author}{\bibfnamefont{O.}~\bibnamefont{Klein}},
  \bibinfo{author}{\bibfnamefont{S.}~\bibnamefont{Donovan}},
  \bibinfo{author}{\bibfnamefont{M.}~\bibnamefont{Dressel}}, \bibnamefont{and}
  \bibinfo{author}{\bibfnamefont{G.}~\bibnamefont{Gr\"uner}},
  \bibinfo{journal}{International Journal of Infrared and Millimeter Waves}
  \textbf{\bibinfo{volume}{14}}, \bibinfo{pages}{2423} (\bibinfo{year}{1993}).

\bibitem[{\citenamefont{Landau and Lifschitz}(1984)}]{LandauBook}
\bibinfo{author}{\bibfnamefont{L.~D.} \bibnamefont{Landau}} \bibnamefont{and}
  \bibinfo{author}{\bibfnamefont{E.~M.} \bibnamefont{Lifschitz}},
  \emph{\bibinfo{title}{Electrodynamics of Continuous Media, Course of
  Theoretical Physics, Vol. 8}} (\bibinfo{publisher}{Pergamon Press},
  \bibinfo{address}{Oxford, UK}, \bibinfo{year}{1984}).

\bibitem[{\citenamefont{Donovan et~al.}(1993)\citenamefont{Donovan, Klein,
  Dressel, Holczer, and Gr\"uner}}]{Gruner2}
\bibinfo{author}{\bibfnamefont{S.}~\bibnamefont{Donovan}},
  \bibinfo{author}{\bibfnamefont{O.}~\bibnamefont{Klein}},
  \bibinfo{author}{\bibfnamefont{M.}~\bibnamefont{Dressel}},
  \bibinfo{author}{\bibfnamefont{K.}~\bibnamefont{Holczer}}, \bibnamefont{and}
  \bibinfo{author}{\bibfnamefont{G.}~\bibnamefont{Gr\"uner}},
  \bibinfo{journal}{International Journal of Infrared and Millimeter Waves}
  \textbf{\bibinfo{volume}{14}}, \bibinfo{pages}{2459} (\bibinfo{year}{1993}).

\bibitem[{\citenamefont{Dressel et~al.}(1993)\citenamefont{Dressel, Klein,
  Donovan, and Gr\"uner}}]{Gruner3}
\bibinfo{author}{\bibfnamefont{M.}~\bibnamefont{Dressel}},
  \bibinfo{author}{\bibfnamefont{O.}~\bibnamefont{Klein}},
  \bibinfo{author}{\bibfnamefont{S.}~\bibnamefont{Donovan}}, \bibnamefont{and}
  \bibinfo{author}{\bibfnamefont{G.}~\bibnamefont{Gr\"uner}},
  \bibinfo{journal}{International Journal of Infrared and Millimeter Waves}
  \textbf{\bibinfo{volume}{14}}, \bibinfo{pages}{2489} (\bibinfo{year}{1993}).

\bibitem[{\citenamefont{Klein et~al.}(1994)\citenamefont{Klein, Nicol, Holczer,
  and Gr\"uner}}]{HolczerPRB}
\bibinfo{author}{\bibfnamefont{O.}~\bibnamefont{Klein}},
  \bibinfo{author}{\bibfnamefont{E.~J.} \bibnamefont{Nicol}},
  \bibinfo{author}{\bibfnamefont{K.}~\bibnamefont{Holczer}}, \bibnamefont{and}
  \bibinfo{author}{\bibfnamefont{G.}~\bibnamefont{Gr\"uner}},
  \bibinfo{journal}{Phys. Rev. B} \textbf{\bibinfo{volume}{50}},
  \bibinfo{pages}{6307} (\bibinfo{year}{1994}).

\bibitem[{\citenamefont{Thurber
  et~al.}(1980{\natexlab{a}})\citenamefont{Thurber, Mattis, Liu, and
  Filliben}}]{BoronDopingResistivity}
\bibinfo{author}{\bibfnamefont{W.~R.} \bibnamefont{Thurber}},
  \bibinfo{author}{\bibfnamefont{R.~L.} \bibnamefont{Mattis}},
  \bibinfo{author}{\bibfnamefont{Y.~M.} \bibnamefont{Liu}}, \bibnamefont{and}
  \bibinfo{author}{\bibfnamefont{J.~J.} \bibnamefont{Filliben}},
  \bibinfo{journal}{Journal of The Electrochemical Society}
  \textbf{\bibinfo{volume}{127}}, \bibinfo{pages}{2291}
  (\bibinfo{year}{1980}{\natexlab{a}}).

\bibitem[{\citenamefont{Thurber
  et~al.}(1980{\natexlab{b}})\citenamefont{Thurber, Mattis, Liu, and
  Filliben}}]{PhosDopingResistivity}
\bibinfo{author}{\bibfnamefont{W.~R.} \bibnamefont{Thurber}},
  \bibinfo{author}{\bibfnamefont{R.~L.} \bibnamefont{Mattis}},
  \bibinfo{author}{\bibfnamefont{Y.~M.} \bibnamefont{Liu}}, \bibnamefont{and}
  \bibinfo{author}{\bibfnamefont{J.~J.} \bibnamefont{Filliben}},
  \bibinfo{journal}{Journal of The Electrochemical Society}
  \textbf{\bibinfo{volume}{127}}, \bibinfo{pages}{1807}
  (\bibinfo{year}{1980}{\natexlab{b}}).

\bibitem[{\citenamefont{Poole}(1996)}]{PooleBook}
\bibinfo{author}{\bibfnamefont{C.~P.} \bibnamefont{Poole}},
  \emph{\bibinfo{title}{Electron Spin Resonance: A Comprehensive Treatise on
  Experimental Techniques}}, Dover Books on Physics (\bibinfo{publisher}{Dover
  Publications}, \bibinfo{year}{1996}), ISBN \bibinfo{isbn}{9780486694443}.

\end{thebibliography}

\appendix
\newpage
\pagebreak
\clearpage

\section{Physical background of the $\mu$-PCD measurement}
In this section, we summarize the most important background knowledge and some supplementary data to the main text.

\begin{figure}[htp]
\begin{center}
\includegraphics[width=0.45\textwidth]{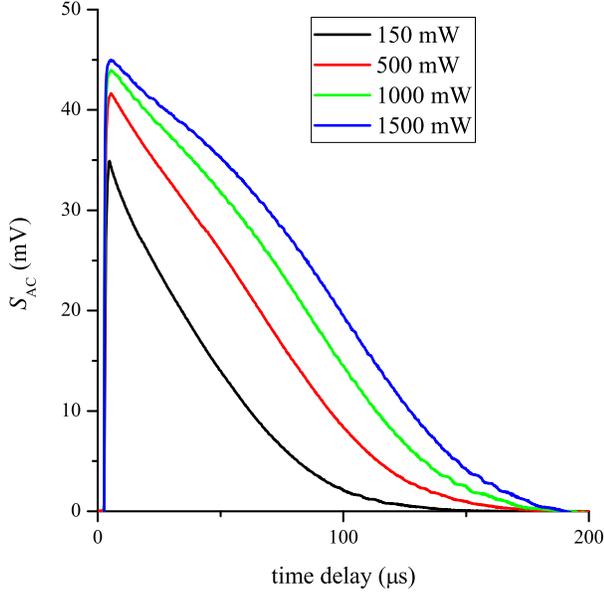}
\caption{$\mu$-PCD traces for a silicon single crystal with $\varrho=19.7\,\Omega\text{cm}$ resistivity for various irradiation powers ($\lambda=527\,\text{nm}$). Note that 1~W average power corresponds to 1 mJ pulse energy.}
\label{Fig:FigSM_DarioAll}
\end{center}
\end{figure}

In Fig.~\ref{Fig:FigSM_DarioAll}. we show the $\mu$-PCD results for a silicon single crystal sample which was detected with the conventional method: a silicon wafer covered entirely a WR90 waveguide. The reflected microwaves were detected from it with and without light illumination. 
In order to calibrate the vertical scale of the $\mu$-PCD traces, it is desired to calibrate the reflected microwave signal voltage by samples with known resistivity. This would enable to obtain the amount of additional charge carriers from the microwave signal. In the following, we denote the reflected signal by $S_{\text{DC}}$ without illumination, and the additional light-induced signal by $S_{\text{AC}}$. We denote the corresponding reflection amplitudes, the $S_{11}$ parameter, as "dark" and "illuminated".

\begin{figure}[htp]
\begin{center}
\includegraphics[width=0.45\textwidth]{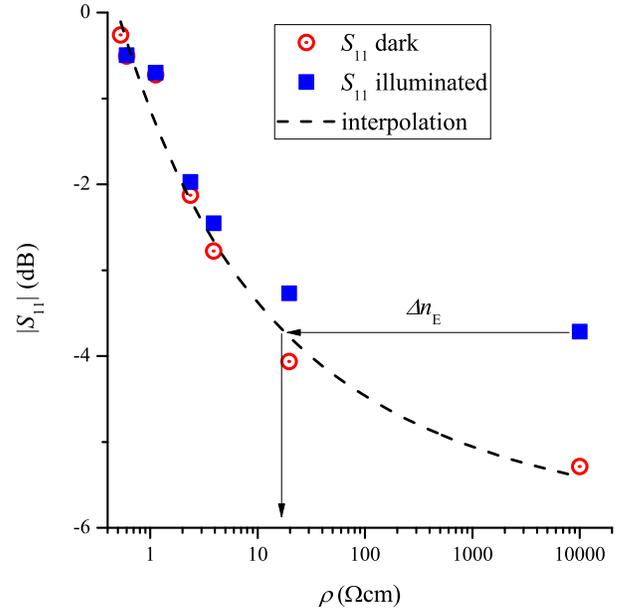}
\caption{Dashed curve is a purely empirical stretched exponential fit as explained in the text. Arrows depict how the illuminated reflection amplitude can be used to determine the sample resistivity under illuminated conditions.}
\label{Fig:FigSM_ACDC}
\end{center}
\end{figure}

Dashed curve is a purely phenomenological interpolation function (i.e. without any theoretical background) which enables to read out the $|S_{11}|$ versus $\varrho$ correspondence. We used $|S_{11}|=-7.08+\exp\left( \frac{1.78}{\varrho^{0.134}}\right)$. 
Clearly, when illuminated, there is an extra reflection due to the metallicity of the sample. The extra reflection can be connected to a modified sample resistivity as arrows depict in the figure. This "illuminated-resisvitity" can be used to determine the amount of light-induced excess charge carrier content from the well-known doping versus resistivity plots \cite{BoronDopingResistivity,PhosDopingResistivity}.

This enabled us to determine the excess charge carrier concentration $\Delta n_{\text{E}}(t)$ for each measurement as a function of time. The latter information is available from the $\mu$-PCD traces which contain the time-dependent $|S_{11}|$.

To complete the analysis, we require the charge carrier recombination time, $\tau_{\text{c}}$, from the $\mu$-PCD traces, also as a function of time. It is known for the light-induced excess charge carriers that the recombination rate depends on the excess charge carrier concentration itself\cite{muPCD_summary1}. This leads to a \emph{time dependence} of $\tau_{\text{c}}$ itself. This can be modelled as $\Delta n_{\text{E}}(t)=A \times \exp \left(-\frac{t}{\tau_{\text{c}}(t)}\right)$.

We obtain: 
\begin{gather}
\tau_{\text{c}}=\left(-\frac{\text{ln}\Delta n_{\text{E}}(t)-\text{ln}A}{t}  \right)^{-1}
\end{gather}
In practice, the $\text{ln}A$ constant subtraction can be performed, which yields the time-dependent $\tau_{\text{c}}$.

\begin{figure}[htp]
\begin{center}
\includegraphics[width=0.45\textwidth]{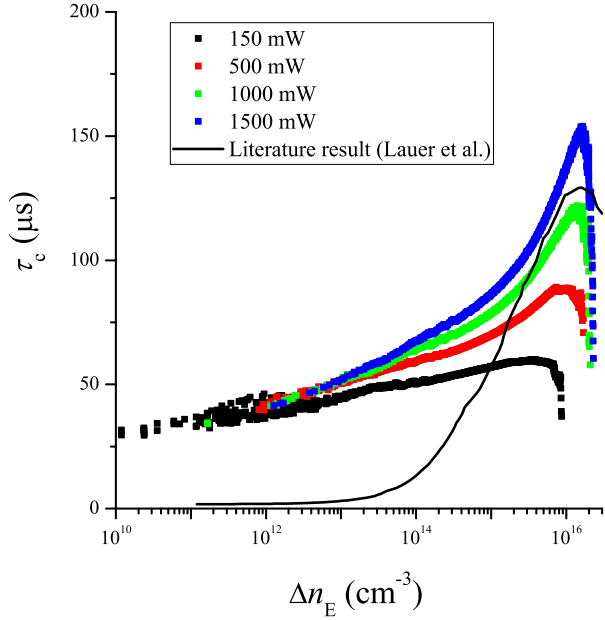}
\caption{$\tau_{\text{c}}$ as a function of the excess charge carrier concentration. Solid curve is a literature data from Ref.~\onlinecite{muPCD_summary1}. }
\label{Fig:FigSM_Delta_n_tau}
\end{center}
\end{figure}

Fig.~\ref{Fig:FigSM_Delta_n_tau}. shows the result of our analysis: namely $\tau_{\text{c}}$ versus $\Delta n_{\text{E}}$ is shown for various exciting laser powers. Ideally, all curves with different powers should fall on one another which is not the case in our data. We speculate that this is due to either heating of the sample or due to charge carrier diffusion. The latter effect influences the microwave reflectivity as the charge carrier concentration is inhomogeneous along the depth profile of the wafer \cite{muPCD_summary1}. Nevertheless, the trends for all curves agree well with the literature data from Ref.~\onlinecite{muPCD_summary1}, especially around the longest $\tau_{\text{c}}$.

The excess charge carrier lifetime is limited by various relaxation rate contributions as follows:
\begin{gather}
\frac{1}{\tau_{\text{c}}}=\frac{1}{\tau_{\text{rad}}}+\frac{1}{\tau_{\text{Auger}}}+\frac{1}{\tau_{\text{SRH}}}
\end{gather}
where $\tau_{\text{rad}}$, $\tau_{\text{Auger}}$, $\tau_{\text{SRH}}$ are the radiative, Auger, Shockley–Read–Hall lifetime contributions, respectively. The radiative lifetime, i.e. electron-hole radiative recombination is significant at high electron-hole concentrations. Similarly, the Auger process (the electron-hole recombination energy is taken away by a free charge carrier) becomes significant for high excess charge carrier concentrations. The Shockley–Read–Hall process occurs due to impurities which form mid-gap states, e.g. Fe and Cr are known to be typical contaminant is silicon. The SRH process probability decreases on higher charge carrier concentrations but importantly it dominates $\frac{1}{\tau_{\text{c}}}$ at low excess charge carrier concentration. Thus measurement of $\frac{1}{\tau_{\text{c}}}$ for low $\Delta n_{\text{E}}$ provides a direct monitoring mean of the impurity content, which is employed in industrial silicon wafer characterization. 

\begin{figure}[htp]
\begin{center}
\includegraphics[width=0.45\textwidth]{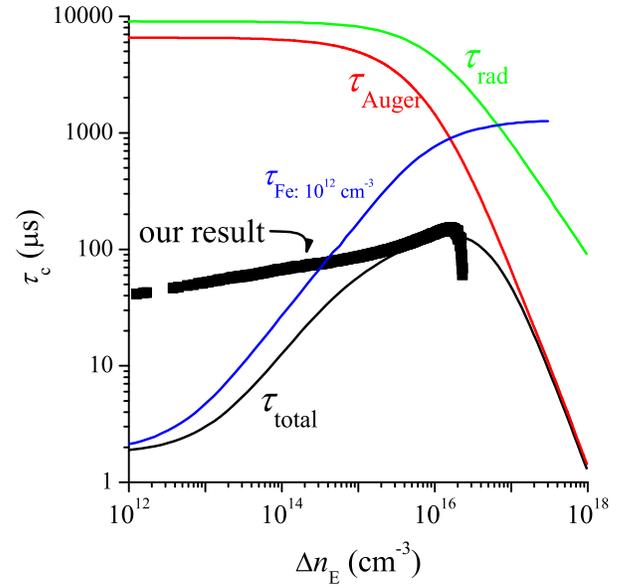}
\caption{Contribution of the different charge recombination processes to the excess charge carrier lifetime after Ref.~\onlinecite{muPCD_summary1}. Symbols are the data points from our measurement at 1500 mW average power. Note the log scale for the charge carrier lifetime.}
\label{Fig:FigSM_tau_literature}
\end{center}
\end{figure}

Fig.~\ref{Fig:FigSM_tau_literature}. shows the contributions from the different excess charge recombination mechanisms and also the resulting total $\tau_{\text{c}}$ for a given Fe impurity content. We also show our data taken at 1500 mW. Note that at the lowest excess charge carrier concentration, our $\tau_{\text{c}}$ value tends to 25 $\mu \text{s}$ which is 10 times longer than the example shown herein, indicating an Fe impurity content (provided Fe is the dominant impurity) below $10^{12}\text{cm}^{-3}$. 

\section{Relation between the generic resonator perturbation and the surface impedance}
\subsection{The case of a cylinder}

Based on Ref.~\onlinecite{LandauBook}, we gave the generic expression for the resonator perturbation for a cylinder with diameter $a$ as:

\begin{gather}
\frac{\Delta f}{f_0}-i\Delta\left(\frac{1}{2Q}\right)=-\gamma \alpha
\label{SM_Landau_cavity_perturbation}
\end{gather}
where $\gamma$ is a sample size dependent constant (also depends on the cavity mode and electromagnetic field distribution). $\Delta f$ is the shift in the resonant frequency and $\Delta\left(\frac{1}{2Q}\right)$ is the change in the resonator bandwidth, BW, (or FWHM) given that $Q=f_0/\text{BW}$ thus $1/2Q=\text{HWHM}/f_0$, where we introduced HWHM (half width at half maximum). The authors of Ref.~\onlinecite{LandauBook} introduced the $\alpha$ polarizability:

\begin{gather}
\alpha=-2\left(1-\frac{2}{a\widetilde{k}}\frac{J_1\left(a\widetilde{k} \right)}{J_0\left(a\widetilde{k} \right)} \right)
\label{SM_Landau_polarizability}
\end{gather}
with $\widetilde{k}=i \omega \sqrt{\mu \epsilon}\sqrt{1-\frac{i}{\omega\epsilon\varrho}}$ being the complex wavenumber of the microwaves inside the material. $J_0$ and $J_1$ are Bessel functions of the first kind.

We then consider the case of finite penetration, i.e. when $\text{Im}\left( \widetilde{k}\right) \to \infty$. Then
\begin{gather}
\lim_{\text{Im}\left( \widetilde{k}\right) \to \infty} \alpha=\lim_{\text{Im}\left( \widetilde{k}\right) \to \infty} -2\left(1-\frac{2}{a \widetilde{k}}\frac{J_1\left(a \widetilde{k}\right)}{J_0\left(a \widetilde{k}\right)}\right)=\\-2+\frac{4i}{a\widetilde{k}}\sim const.+i Z_{\text{s}}.
\label{SM_derivation}
\end{gather}
The relation between the surface impedance and the wave vector is as follows: $Z_{\text{s}}=Z_0/\widetilde{n}$ and $\widetilde{k}=\omega\widetilde{n}/c$ (with $c$ being the speed of light), which yields: $Z_{\text{s}}=Z_0\omega/\widetilde{k}c=Z_0/\widetilde{k}\lambda_0$, where $\lambda_0$ is the wavelength of the electromagnetic wave in vacuum.

We have also used the identity: 
\begin{gather}
\lim_{y \to \infty} \frac{J_1(x+i y)}{J_0(x+i y)}=i
\end{gather}

The $const.$ in Eq.~\eqref{SM_derivation} expresses the fact that the resonator shift is referenced to a \emph{perfect} conductor ($\sigma=\infty$), i.e. one which expels all the electromagnetic fields. This derivation leads us to the well-known formula for the resonator perturbation, which contains the surface impedance\cite{PozarBook,chen2004microwave,Gruner1,Gruner2,Gruner3}:

\begin{gather}
\frac{\Delta f}{f_0}-i\Delta\left(\frac{1}{2Q}\right)=-i \nu Z_{\text{s}}
\label{SM_surface_imp_cavity_perturbation}
\end{gather}
where $\nu$ is a geometry factor (not dimensionless) that is proportional to the sample surface to the surface of the cavity but it also depends on the resonator mode. 

We also note that the shown Re and Im values of $\alpha$ can be obtained to match one another when these are shifted by a constant for the case of $\sigma \rightarrow \infty$.

\subsection{The case of a sphere}

Similarly as before, we can calculate the polarizability of sphere samples from the Helmholtz equation, then we can obtain $\Delta f$ and $\Delta \left(\frac{1}{2Q}\right)$ from Eq.~\eqref{SM_Landau_cavity_perturbation}. The polarizability of a sphere sample with diameter $a$ is:
\begin{equation}
	\alpha=-\frac{3}{2}\left(1-\frac{3}{a^2\widetilde{k}^2}+\frac{3}{a\widetilde{k}}\cot\left(a\widetilde{k}\right)\right),
\end{equation}
where the complex wavenumber is the same as before.

In the case of finite penetration:
\begin{gather}
	\lim_{\text{Im}\left( \widetilde{k}\right) \to \infty} \alpha=\lim_{\text{Im}\left( \widetilde{k}\right) \to \infty} -\frac{3}{2}\left(1-\frac{3}{a^2\widetilde{k}^2}+\frac{3}{a \widetilde{k}}\cot\left(a \widetilde{k}\right)\right)=\\-\frac{3}{2}+\frac{9i}{2a\widetilde{k}}\sim const.+i Z_{\text{s}},
	\label{SM_derivation_sphere}
\end{gather}
where we use the identity:
\begin{gather}
	\lim_{y\rightarrow\infty}\cot\left(x+iy\right)=-i.
\end{gather}

Note that, the $const.$ terms are different in Eq.~\eqref{SM_derivation} and Eq.~\eqref{SM_derivation_sphere} due to the different sample geometry.

\section{The effect of the dielectric constant on the cavity perturbation}

\begin{figure}[htp]
\begin{center}
\includegraphics[width=0.45\textwidth]{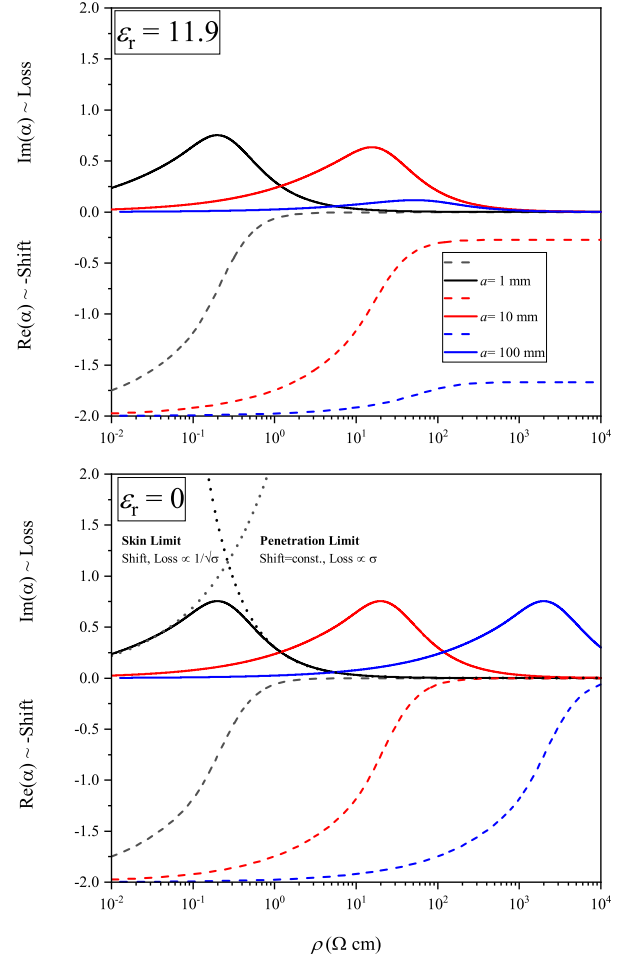}
\caption{Variation of the $\alpha$ parameter after Ref.~\onlinecite{LandauBook} for a realistic case of silicon ($\epsilon_{\text{r}}=11.9$) and for a good metal, when the displacement effects are neglected ($\epsilon_{\text{r}}=0$). The lower panel also shows the asymptotic behaviors (dotted curves) for the skin limit: Loss~$\propto 1/\sqrt{\sigma}$, and for the penetration limit: Loss~$\propto \sigma$ behaviors.}
\label{Fig:SM_Landau_formula_epsilon_r}
\end{center}
\end{figure}

In Fig.~\ref{Fig:SM_Landau_formula_epsilon_r}., we show the effect of a finite $\epsilon_{\text{r}}$ for the resonator shift and loss as calculated for a cylinder with varying diameter. Note that in the absence of displacement current related effects ($\sigma\gg \epsilon \omega$), both the loss and resonator shift terms have the same magnitude. The figure also shows the asymptotic behaviors (doted curves) for the skin limit: Loss $\propto 1/\sqrt{\sigma}$, and for the penetration limit: Loss $\propto \sigma$ behaviors. When shifted by 2, the shift value matches exactly the loss for the $\epsilon_{\text{r}}=0$ case in the skin-limit.

\section{A lumped circuit model calculation of the resonator enhancement effect}

We first consider a conventional RLC circuit whose frequency-dependent impedance reads near resonance ($\omega_0=1/\sqrt{L C}$):
\begin{gather}
Z(\omega)_{\text{unmatched}}\approx R+i2R Q_0 \frac{\omega-\omega_0}{\omega_0},
\label{unmatchedRLC}
\end{gather}
where the unloaded quality factor reads $Q_0=L\omega_0/R$.

\begin{figure}[htb]
\begin{center}
\includegraphics[width=.5\linewidth]{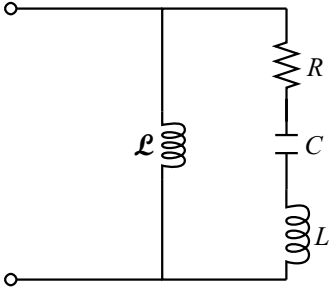}
\caption{RLC model of a coupled microwave cavity. Note that the inductor $\mathcal{L}$ models the matching element.}
\label{matched_cavity}
\end{center}
\end{figure}

We then consider an RLC circuit whose impedance is \emph{matched} to the wave impedance of the waveguide, $Z_0$. One can model the matching of microwave resonators by the lumped circuit model in Fig.~\ref{matched_cavity} after Refs.~\onlinecite{PooleBook,PozarBook}. 
The frequency dependent impedance of such a resonator near the resonance, $\omega\approx \omega_0$ reads:
\begin{gather}
Z(\omega)_{\text{matched}}\approx Z_0\pm i2Z_0 Q \frac{\omega-\omega_0}{\omega_0},
\label{Near_resonance_impedance}
\end{gather}
where $Q=Q_0/2$ is the quality factor of a critically coupled resonator.

The minus sign difference stems from the type of matching element; the sign is + for a capacitive and - for inductive matching (such as that in the figure). Clearly, the difference between Eq.~\eqref{Near_resonance_impedance} and Eq.~\eqref{unmatchedRLC}. is that on resonance its impedance is \emph{transformed} from the original $R$ to $Z_0$. It is less well known that this transformation property is the principal underlying factor why one uses resonators at all, and how the presence of resonators essentially magnify the sensitivity of material properties measurements. 

To demonstrate this, we explicitly express the dependence of the matched resonator parameters on the circuit parameters and then we consider a small perturbation to $R$. The perturbation can be thought of as a small extra absorption in the circuit due to the presence of a sample (or eddy current). It can be shown without the loss of generality that similar conclusion can be drawn when the inductivity in the original circuit is perturbed, e.g. by a piece of a magnetic sample such as that using magnetic resonance.

Ref.~\onlinecite{PozarBook} derives that for the above circuit the resonance and impedance matching conditions are:

\begin{equation}
     \mathcal{L}^{2} \omega_{0}^{2} = R Z_{0},
 \end{equation}
 where $\omega_{0} = 1/\sqrt{C\left(L+\mathcal{L}\right)}$ holds.

Clearly, this equation sets the value of $\mathcal{L}$. In the high $Q$ limit, $Z_0\gg R$, thus $\mathcal{L}\ll L$, it thus also shows that the resonance frequency is only slightly shifted with respect to $\omega_0=1/\sqrt{LC}$.

We then consider the sensitivity of the circuit return impedance (or $Z(\omega)$) with respect to $R$. This is obtained from the change in the corresponding impedances as a function of a small perturbation in $R$: $R\rightarrow R+\Delta R$. We obtain:
\begin{equation}
  \Delta \mathrm{Re} Z_{\text{unmatched}}\left(\Delta R \right) = \left.\frac{\partial \mathrm{Re}\,Z_{\text{unmatched}}}{\partial R}\right|_{\omega = \omega_{0}}\Delta R =  \Delta R.
	\label{sensitivity_unmatched}
\end{equation}
where we used that for an unmatched circuit, such as that described by Eq.~\eqref{unmatchedRLC}, the following derivative reads:
\begin{equation}
  \left.\frac{\partial \mathrm{Re}\,Z_{\text{unmatched}}}{\partial R} \right|_{\omega = \omega_{0}} =  1.
	\label{derivative_unmatched}
\end{equation}

The sensitivity of the real part impedance of a matched circuit is on the other hand:
\begin{equation}
  \Delta \mathrm{Re} Z_{\text{matched}}\left(\Delta R \right) = \left.\frac{\partial \mathrm{Re}\,Z_{\text{matched}}}{\partial R} \right|_{\omega = \omega_{0}}\Delta R = \pm \frac{Z_{0}}{R} \Delta R.
	\label{sensitivity_matched}
\end{equation}
where we used that for the impedance of the matched circuit described by Eq.~\eqref{matched_cavity} the derivative reads:
\begin{equation}
  \left.\frac{\partial \mathrm{Re}\,Z_{\text{matched}}}{\partial R} \right|_{\omega = \omega_{0}} = \pm \frac{Z_{0}}{R}.
	\label{derivative_matched}
\end{equation}

We note that the corresponding first order derivatives for the imaginary parts vanish near resonance for both cases. The striking fact about Eqs.~\eqref{sensitivity_unmatched} and \eqref{sensitivity_matched} is that the matched circuit appears to act as an impedance transformer by $Z_0/R$. We also note that other cases of the resonator perturbation can be similarly considered. E.g. when the resonator is perturbed by a magnetic material, its effect can be taken into account as a change in $L$, as: $L\rightarrow L(1+
\widetilde{\chi})$, where $\widetilde{\chi}=\chi'+i \chi''$ is the (complex) magnetic susceptibility. The $\chi''$ acts as if $R$ was perturbed by $L\omega_0\chi''$. Thus the above argument applies and the sensitivity for this perturbation reads and its effect is amplified by $Z_0/R$. 

The real part, $\chi'$ perturbes $L$ by $\Delta L=L\omega_0\chi'$, which has an effect on the imaginary part of $Z(\omega)$. This case: 
\begin{equation}
  \Delta \mathrm{Im} Z_{\text{unmatched}}\left(\Delta L \right) = \frac{\partial \mathrm{Im}\,Z_{\text{unmatched}}}{\partial L}\Delta L =  2 L\omega_0\chi' \frac{\omega-\omega_0}{\omega_0}.
	\label{L_sensitivity_unmatched}
\end{equation}

For the matched case, we obtain:
\begin{equation}
  \Delta \mathrm{Im} Z_{\text{matched}}\left(\Delta L \right) = \frac{\partial \mathrm{Im}\,Z_{\text{matched}}}{\partial L}\Delta L =  \frac{Z_0}{R} 2 L\omega_0\chi' \frac{\omega-\omega_0}{\omega_0}.
	\label{L_sensitivity_matched}
\end{equation}

The $Z_0/R$ enhancement factor is often mistaken by an enhancement effect by $Q$ (or $Q_0$), the reason being that for most resonators $Z_0 \approx L\omega_0$ holds thus $Z_0/R\approx L \omega_0/R=Q_0$. The $L\omega_0\approx Z_0$ can be motivated for a waveguide and a corresponding resonator: a fundamental mode rectangular resonator with a mode of TE101, which is made out of a half wavelength section of a TE10 cylindrical waveguide. For both the TE101 cavity and for the $\lambda/2$ TE10 section, the inductivity is $L$, and capacitance is $C$. Given that $Z_0=\sqrt{L/C}$ and $\omega_0=1/\sqrt{L C}$, we get exactly $Z_0=L\omega_0$. Similar arguments hold for other types of resonators such as e.g. a $\lambda/2$ resonator made of a coplanar waveguide \cite{PozarBook}.

We finally show that one observes a similar up-transformation (i.e. enhancement) effect for the reflection coefficient. Again, we consider the case of the unmatched and matched circuits described by Eqs.~\eqref{sensitivity_unmatched} and \eqref{sensitivity_matched}, respectively. The reflection coefficients read near resonance (assuming $Z_0 \gg R$ due to the large $Q$):

 \begin{equation}
     \Gamma_{\text{unmatched}} = \frac{R-Z_{0}}{R+Z_{0}} \approx 1 - 2 \frac{R}{Z_{0}},\\
     \Gamma_{\text{matched}} = \frac{Z-Z_{0}}{Z+Z_{0}} =0,\\
\end{equation}
 thus the reflection coefficient is close to 1 for the unmatched case which is often disadvantageous, whereas the matched case represents a \emph{null measurement}.

The corresponding derivatives read:
\begin{equation}
   \left.\frac{\partial Z_{\text{unmatched}}}{\partial R} \right|_{\omega = \omega_{0}} \approx - \frac{2}{Z_{0}},\\
   \left.\frac{\partial Z_{\text{matched}}}{\partial R} \right|_{\omega = \omega_{0}} \approx \pm \frac{1}{2R}.
\end{equation}
Therefore the sensitivity of the reflection coefficient is enhanced by $Z_0/4R$ for the case of the matched circuit as compared to the unmatched case.

\section{The resonator advantage over a conventional reflection setup}

The above discussion is valid for a conventional reflection setup, where the reflected RF voltage is detected with a continuous wave irradiation. As it was shown, the reflectometry method is more sensitive for a matched resonator that for a simple unmatched circuit.

It is also worth discussing the case when the resonator parameters, the frequency shift ($\Delta f$) and the $Q$ factor change ($\Delta\left(\frac{1}{2Q}\right)$), are measured directly. Without the loss of generality, we consider the case of a magnetic sample, whose effect can be well demonstrated. The magnetic sample with a complex susceptibility of $\widetilde{\chi}$ perturbs a solenoid of an RF circuit as: $L\rightarrow L(1+\eta\widetilde{\chi})$, where $\eta$ is the filling factor. Such a sample perturbs the resonator parameters as\cite{PooleBook,chen2004microwave}: 
\begin{gather}
\frac{\Delta f}{f_0}-i\Delta\left(\frac{1}{2Q}\right)=-\eta \widetilde{\chi}
\label{Chi_perturbation}
\end{gather}

In the following, we describe the error of the $\eta\widetilde{\chi}$ measurement for the non-resonant and resonant cases. In the conventional reflectometry technique, it is obtained from the reflection coefficient, $\Gamma$. We consider a waveguide with wave impedance $Z_0$, which is terminated by an inductor with inductance $L$. We then introduce the empty reflection coefficient (i.e. without the sample), $\Gamma_{\text{empty}}$, and that with the sample, $\Gamma_{\text{sample}}$. This gives:
\begin{gather}
\eta \widetilde{\chi}\approx\frac{iL\omega+Z_0}{iL\omega}\left(\Gamma_{\text{sample}}-\Gamma_{\text{empty}} \right),
\end{gather}
where we retained leading order terms in $\eta \widetilde{\chi}$ only.

We introduce the standard error of the respective measurements as $\sigma\left(.\right)$. Error propagation dictates that 
\begin{gather}
\sigma\left(\eta\left|\widetilde{\chi}\right|\right)_{\text{non-resonant}}=2\frac{\sigma\left(\Gamma\right)}{\Gamma}.
\label{error_in_non_resonant}
\end{gather}
the $\left| \right|$ notation is employed as $\widetilde{\chi}$ is a complex quantity.
Our experience with the conventional reflectometry setup using VNAs shows that the quantity on the right hand side is about $10^{-3}..10^{-4}$, which fixes the attainable accuracy of the susceptibility measurement.

On the other hand, we showed previously \cite{GyureRSI,GyureGaramiRSI} that the standard error of $\frac{\Delta f}{f_0}$ can be expressed as: 
\begin{gather}
\frac{\sigma\left(\Delta f\right)}{f_0}=\frac{\sigma\left(\Delta f\right)}{\text{BW}}\frac{\text{BW}}{f_0}=\frac{1}{Q}\frac{\sigma\left(\Delta f\right)}{\text{BW}}. 
\label{error_in_resonant}
\end{gather}
where we introduced the resonator bandwidth, BW, which is related to the $Q$ as $f/\text{BW}=Q$. We assumed that $f_0$ is error free as it is a dividing constant. We showed in Refs. \onlinecite{GyureRSI,GyureGaramiRSI} that the quantity $\frac{\sigma\left(\Delta f\right)}{\text{BW}}$ is typically $10^{-3}..10^{-4}$. Clearly, a comparison between Eqs. \eqref{error_in_non_resonant} and \eqref{error_in_resonant} yields that again, the enhancement in the accuracy of the resonator based measurement is $Q$-fold.

The enhancement can be obtained similarly for the $Q$ factor change, by realizing that the error of the shift measurement is the same as the measurement of the BW as it was shown in Refs. \onlinecite{GyureRSI,GyureGaramiRSI}.

\end{document}